\documentclass[11pt]{article}
\usepackage{amsmath}
\usepackage{color}
\usepackage{amssymb}
\usepackage{graphicx,psfrag,epsf}
\usepackage{enumerate}
\usepackage{url,appendix} 
\usepackage{longtable}
\usepackage{framed}
\usepackage{rotating}
\usepackage{caption}
\usepackage{pbox}
\usepackage{booktabs}
\usepackage{booktabs}
\usepackage{rotating}

\definecolor{marro}{rgb}{0.6,0,0}

\newcommand{\Prob}{\mathbb{P}}
\newcommand{\ignore}[1]{}
\newcommand{\rowgroup}[1]{\hspace{0em}\textbf{#1}}
\newcommand*{\MyIndent}{\hspace*{0.5cm}}

\begin{document}
\bibliographystyle{plain}
\title{Bayesian Propensity Scores for High-Dimensional Causal Inference: A Comparison of Drug-Eluting to Bare-Metal Coronary Stents}

\author{Jacob V Spertus$^{1}$ and Sharon-Lise T Normand$^{1,2}$ \\
	\small{1: Department of Health Care Policy, Harvard Medical School}\\ \small{2: Department of Biostatistics, Harvard TH Chan School of Public Health}}
\date{}
\maketitle 

\begin{abstract}
{High-dimensional data can be useful for causal inference by providing many confounders that may bolster the plausibility of the ignorability assumption. Propensity score methods are powerful tools for causal inference, are popular in health care research, and are particularly useful for high-dimensional data. Recent interest has surrounded a Bayesian formulation of these methods in order to flexibly estimate propensity scores and summarize posterior quantities while incorporating variance from the (potentially high-dimensional) treatment model. We discuss methods for Bayesian propensity score analysis of binary treatments, focusing on modern methods for high-dimensional Bayesian regression and the propagation of uncertainty from the treatment regression. We introduce a novel and simple estimator for the average treatment effect that capitalizes on conjugancy of the beta and binomial distributions. Through simulations, we show the utility of horseshoe priors and Bayesian additive regression trees paired with our new estimator, while demonstrating the importance of including variance from the treatment and outcome models. Cardiac stent data with almost 500 confounders and 9000 patients illustrate approaches and compare among existing frequentist alternatives.}\\

\end{abstract}

\section{Introduction}
\label{sec:intro}

\indent

Comparative effectiveness research examines which health care interventions work best for which patients. These questions are typically causal in nature and frequently informed by randomized clinical trials (RCTs). When experimental evidence is unavailable, causal inferences may be drawn from observational data such as information found in health care databases. Propensity score approaches that condition on treatment assignment probabilities are often used to adjust for confounding and attain unbiased causal effects provided key assumptions are met. They have a central role in applied causal inference for inferring effectiveness of interventions in usual care settings or in regulatory settings for clearance of medical devices \cite{fda2013}.  One key feature to the validity of findings is that propensity score approaches build treatment models without reference to outcome information in analogy to randomized trials, an {\em outcome free design}.  This feature is vital for high-dimensional causal inference problems where many potential confounders are available \cite{rubin2008}\cite{austin2015}\cite{schneeweiss2009}.

Recent work has attempted to bring propensity scores into the domain of Bayesian inference. It is difficult to reconcile the Bayesian paradigm, in which all parameters of interest are jointly estimated, with propensity scores, in which outcome and treatment models are explicitly separated. Nevertheless, Bayesian methods offer their own suite of tools for model building, provide intuitive ways to present posterior summaries of parameters, and propagate uncertainty from all stages of model fitting  \cite{mccandless2009}\cite{kaplan2012}\cite{zigler2014}\cite{saarela2015}\cite{graham2016}. For these reasons, Bayesian propensity scores remain an appealing concept and an exciting area for methodological advancement. The propagation of uncertainty is a particularly important feature for high-dimensional causal inference, as the treatment model becomes the main source of variance.
 
Our work is motivated by a comparison of the effectiveness of two treatment alternatives for percutaneous coronary intervention (PCI) in a cohort of patients treated in Massachusetts non-federal hospitals. These interventions are standard care for clearing blockages in the blood vessels supplying the heart and frequently include the installation of a coronary stent to keep the treated artery clear and supported. We focus on two classes of coronary stents available in the U.S. market: bare-metal (BMS) which were first approved in the U.S. in the 1990s and drug-eluting (DES) approved in the U.S. in 2003. Many randomized clinical trials, meta-analyses, and observational studies have compared the effectiveness of DESs and BMSs for patients with a range of conditions \cite{venkitachalam2011}\cite{mauri_circ2008}\cite{bonaa2016}. Our main data source is the Mass-DAC clinical registry, a state mandated database harvesting clinical information for all PCIs performed in adults (age $\geq$ 18 years) in all non-federal Massachusetts' hospitals annually. We sought to estimate the causal effect of DES compared to BMS on repeat target-vessel revascularization (TVR) and on all-cause mortality within 1 year of the initial implant. Our data, examining stents implanted in 2011, includes about 9,000 patients undergoing coronary stenting and approximately 500 confounders. A small subset of these confounders and outcomes appears in table \ref{small_confounder_table}. Generally, BMS patients are sicker and older than DES patients, confounding a naive estimate of causal effect based on raw outcomes. 

\begin{table}[!h]
	\centering
	\caption{\textit{Prevalence or mean (SD) of selected confounders and outcomes in Mass-DAC data. TVR = target vessel revascularization; STEMI = ST-elevated myocardial infarction; LVSD = left ventricular systolic dysfunction.}}
	\begin{tabular}{lrrr}
		&& \multicolumn{2}{c}{Stent Type} \\ \cline{3-4}
		Characteristic & Overall & Bare Metal & Drug Eluting \\ 
		\hline
		\textbf{Outcomes} &\\
		1-Year Mortality, \% & 5.7 & 10.2 & 3.3 \\ 
		1-Year TVR, \% & 7.4 & 9.0 & 6.5 \\ \hline
		\textbf{Confounders} & \\
		Mean Age (SD), yrs & 64.7 (12.5) & 66.4 (11.7) & 63.7 (13.5) \\ 
		STEMI, \% & 24.4 & 35.7 & 18.2 \\ 
		Cardiomyopathy or LVSD, \% & 9.4 & 11.1 & 8.4 \\ 
		Emergent Status, \% & 26.7 & 38.3 & 20.3 \\ 
		Cardiogenic Shock, \% & 1.8 & 3.8 & 0.8 \\ 
		\hline
	\end{tabular}
	\label{small_confounder_table}
\end{table}

We propose a new method of Bayesian propensity score analysis for causal inference, with emphasis on using high-dimensional data to meet the ignorability assumption - the assumption that treatment assignment does not depend on the potential outcomes. Our method estimates a difference in outcome proportions in a weighted pseudo-population, sharing links to frequentist inverse probability weighting (IPW) as well as the weighted likelihood approach of \cite{saarela2015}. By focusing on a Bayesian framework, we allow for intuitive and flexible approaches to model building with many variables while propagating uncertainty in the treatment model. We do not offer new theory on Bayesian propensity scores, referring the interested reader to the extensive discussion in \cite{saarela2015}, and its accompanying critiques. We assess our approach on the basis of finite sample frequentist characteristics such as consistency, efficiency, and coverage through simulation studies.
 

We next briefly review the goals and assumptions of causal inference in section 2. Section 3 provides background on propensity scores as a causal inference tool, introduces Bayesian methods for regression with high-dimensional data, and discusses recent work in developing a Bayesian framework for propensity scores. In section 4 we provide technical details and propose a two-step propensity score weighting procedure using Bayesian computation. Finite sample performance obtained through a simulation study of our methods and competing existing approaches is described in section 5. We return to the coronary stent problem in section 6.  Throughout we assume a binary treatment, referring to one arm as treatment and the other arm generically as control. We label the ``treatment model" as the model characterizing the relationship between confounders and treatment, and the ``outcome model" as the model characterizing the relationship between treatment and outcome, adjusting for confounders.

\section{Assumptions for Causal Inference}

\indent 

We make three assumptions necessary for causal inference: positivity, stable unit treatment value assignment (SUTVA), and ignorability. SUTVA encodes the assumptions that observations do not interfere with one another and that the nature of treatment does not vary across individuals. Positivity refers to  the assumption that each subject has a chance for assignment to either treatment group. It is difficult to test for positivity violations in high-dimensions, as each subject is typically uniquely defined by some combination of their covariates. We do assume positivity is met in our data and refer the reader to our references for more information \cite{petersen2010}. Our focus is on meeting the ignorability assumption by exploiting high-dimensional data.

\subsection{Ignorability}

\indent

For a subject $i$, let $X_i$ denote binary treatment, $Y_{1i}$ and $Y_{0i}$ potential outcomes under treatment and control respectively, and $\boldsymbol C_i$ a vector of pre-treatment confounders. We are interested in the treatment effect, $\mathbb{E}(Y_{0i}) - \mathbb{E}(Y_{1i})$. However we can never observe both potential outcomes for a subjects. We can use observational data to estimate the treatment effect if we are willing to assume that conditional on observed pre-treatment confounders, treatment is independent of potential outcomes across the population \cite{rosenbaum1983}:
$$({Y_{0i}},  {Y_{1i}}) \perp {X_i} | \boldsymbol{C}_i.$$
Coupled with positivity, ignorability permits estimation of the average treatment effect (ATE) defined here as the difference in marginal event rates under each treatment across the population:
$$\Delta = \mathbb{E}_\mathbf{C}[\mathbb{E}(\boldsymbol Y | \boldsymbol X = 1, \mathbf{C}) - \mathbb{E}(\boldsymbol Y | \boldsymbol X = 0, \mathbf{C})].$$
If an important confounder is excluded from $\mathbf C$, then the ignorability assumption is not met and our estimate of $\Delta$ will be biased. Thus in the observational setting, inclusion of important confounders is essential to generating unbiased causal estimates.

\section{Causal Inference and Bayesian Methods}

\subsection{Propensity Scores}

\indent

For binary treatment the propensity score for subject $i$ is defined as $\pi(\boldsymbol{C}_i) = \Prob(X_i = 1 | \boldsymbol C_i)$, the probability of receiving treatment given pre-treatment confounders. In their seminal paper, \cite{rosenbaum1983} define the vector $\boldsymbol \pi(\mathbf{C}) = [\pi(\boldsymbol{C}_1)...\pi(\boldsymbol{C}_n)]$ as a ``balancing score'' and show that conditioning on it is sufficient to estimate unbiased causal effects, i.e. $(\boldsymbol{Y_0,  Y_1}) \perp \boldsymbol{X} | \boldsymbol \pi(\mathbf{C})$. This fact is particularly useful in the high-dimensional setting because we only need to condition on the one-dimensional propensity score vector, $\boldsymbol \pi(\mathbf{C})$, rather than the entire matrix of confounders, $\mathbf{C}$ \cite{rosenbaum1983}. However, in the absence of randomization, the propensity scores must still be estimated from a regression of confounders on treatment and a first step involves reducing the dimension of the confounders via variable selection or by some method of regularization. Focusing on the treatment model only, an outcome free design, is preferable over estimates based on outcome regression adjustment because it reduces the risk of selective inference \cite{rubin2008}. The user can design a good propensity score model by assessing balance or overlap among confounders between treatment groups and tweaking the propensity score model if necessary \cite{austin2015}. Because this approach reduces investigator bias and shares important connections to randomized trials, propensity score methods are recommended for use in observational studies by important regulators and decision-makers \cite{fda2013}. 

\subsection{Bayesian Propensity Scores} \label{sec:bayesian_propensity_scores}

\indent

Several researchers have justified and implemented propensity scores in a fully Bayesian framework by jointly modeling the outcome and treatment assignment mechanism \cite{mccandless2009}\cite{kaplan2012}\cite{zigler2014}. In addition to adhering to the likelihood principle, an advantage of these approaches is that instrumental variables can be screened out by the use of carefully constructed priors that select variables associated with both treatment and outcome \cite{zigler2014}. Nonetheless, strong critiques have been raised involving three key aspects of fully Bayesian propensity score methods. First, when the outcome model parameters are \textit{a-priori} independent of the treatment model, the propensity scores play no role in Bayesian inference. Second, if a-priori treatment and outcome parameters are dependent then joint modeling of outcome and treatment change the treatment balancing property of the propensity score.  Finally, joint modeling can be very sensitive to model misspecification, leading to poor calibration and coverage \cite{zigler2013}\cite{zigler2014}\cite{saarela2015}\cite{graham2016}.  

We instead pursue a method along the lines of \cite{kaplan2012} and \cite{saarela2015}, who developed Bayesian computational approaches by estimating the treatment and outcome models in two stages. Due largely to the first criticism above, such a method is not widely accepted as formally Bayesian \cite{saarela2015}. Our treatment and outcome models are separately Bayesian, but the way they are integrated is not. Thus our estimator must ultimately be assessed in terms of frequentist characteristics like coverage and consistency. Furthermore, because we maintain outcome free design, there is inherently no way to automatically screen instrumental variables and, as in a typical propensity score analysis, we rely on background knowledge to exclude instruments \cite{austin2015}. From a pragmatic viewpoint, we maintain the outcome free design feature while allowing the use of Bayesian modeling strategies (Bayesian algorithms, priors, posterior sampling, etc) for the propensity score model. Additionally, we propagate uncertainty from both the propensity score and outcome model which improves frequentist properties of causal estimators in high-dimensions \cite{zigler2014}. As a further advantage, we generate the full posterior density for the causal parameter of interest, allowing us to present results more flexibly than we could relying on point estimates \cite{graham2016}.     

\section{Proposed Approaches}

\indent 

We propose new approaches for both the treatment regression and outcome model of a two-step Bayesian propensity score analysis. For the treatment regression, we estimate Bayesian logistic models regularized by weakly informative or sparsity inducing priors, as well as fully non-parametric Bayesian additive regression trees. For the outcome model, we sample from separate beta posterior distributions whose arguments are based on a weighted number of counts observed in the samples.

\subsection{Regularization}

\indent 

Regularization is a popular technique in high-dimensional data analysis. Suppose we wish to estimate a logistic regression model to predict a binary treatment $\boldsymbol X$ from confounders $\mathbf{C}$ with coefficients $\boldsymbol{\beta} = (\beta_0, \beta_1, ... \beta_p)$:
\begin{align}
X_i \sim \mbox{Bernoulli} \bigg \{ \mbox{logit}^{-1}(\beta_0 + \sum_{j=1}^p \beta_j C_{ij}) \bigg \}. \label{logistic_regression}
\end{align}

\noindent
If $p$, the dimension of $\mathbf{C}$, is large, the parameter vector $\boldsymbol{\beta}$ becomes highly variable and the model tends to ``over-fit" the data. Regularization trades a bit of bias for a relatively large efficiency gain when estimating $\boldsymbol{\beta}$. This controls overfitting and can reduce the mean squared error of estimates. From a Bayesian perspective, regularization amounts to placing informative prior distributions on parameters, often centered at zero.  Commonly used regularizing distributions include normal, double-exponential (Bayesian lasso), and Student-t distributions \cite{gelman2014}. Choosing to exclude variable $j$ based on subject matter knowledge is equivalent to prior certainty that $\beta_j = 0$.
 
Specification of priors can be very intuitive. For example, in a logistic regression we might assume with 95\% prior certainty that a binary confounder will not have a log-odds effect greater than 5 in magnitude, which is quite large in our motivating example. A zero-centered Student-t distribution with 3 degrees of freedom and scale 2.5 accomplishes this, while heavy tails accommodate larger effect sizes if the data warrant. These are called ``weakly informative priors" \cite{gelman2014}. With more complex prior specifications some interpretability may be sacrificed for improved performance in high-dimensional settings. Ideal regularization priors separate confounders from noisy variables, aggressively shrinking covariates towards zero to reduce variance while allowing the coefficients of true confounders to remain relatively unbiased. The horseshoe prior is a Bayesian regularization tool designed for this purpose \cite{carvahlo2010}. It has both a very high density near zero and fat tails that do not over-shrink true signals. The horseshoe prior can be expressed as a scale mixture of normal distributions: 
\begin{align*}
\beta_j &\sim \mathcal{N}(0, \lambda_j^2 \tau^2) ~\mbox{where}~ \lambda_j \sim \mbox{Cauchy}^+(0, 1)  ~~\mbox{and}~~ \tau \sim \mbox{Cauchy}^+(0,1).
\end{align*}      
\noindent
The parameter $\tau$ acts as a global scale parameter and controls the overall degree of shrinkage; $\lambda_j$ acts as a local scale parameter for each coefficient, allowing confounder coefficients to be quite large while shrinking noisy coefficients towards zero. Weakly informative half-Cauchy priors on $\lambda_j$ and $\tau$ imply a fully Bayesian specification \cite{gelman2014}. With default specifications, horseshoe priors outperform the cross-validated lasso in sparse prediction problems and provide fully Bayesian inference \cite{carvahlo2010}. 

Regularization in causal inference has garnered much attention, most of which is predicated on outcome modeling. For a thorough review of recent developments in high-dimensional causal inference see \cite{hahn2016}. Recent highlights in the frequentist literature include \cite{ghosh2015} that proposed lasso regularization for causal inference based on outcome regression, and \cite{farrell2015} that developed theory for doubly-robust estimation using the lasso. On the Bayesian side, \cite{wang2012} proposed an algorithm using discrete priors with linked treatment and outcome parameters to select variables likely to be confounders. Hahn et al \cite{hahn2016} used horseshoe priors with a carefully chosen parameterization to estimate a treatment effect in a regularized Bayesian outcome regression and \cite{zigler2014} proposed a joint Bayesian propensity score approach with discrete regularizing priors. 

The downside to regularizing priors paired with logistic regression is that the user must pre-specify the correct model form including interactions and non-linearities. This can be particularly challenging when many potential confounders are available.

\subsection{Bayesian Additive Regression Trees (BART)}

\indent

BART is a nonparametric modeling technique that translates decision tree-based ensemble methods, such as random forests, to a Bayesian framework. Such approaches are especially desirable when the form of the model is unknown. \cite{chipman2010} provide a thorough introduction to the method. Broadly speaking, BART is a sum-of-trees model where prior distributions are placed over the parameters including tree depth, splitting variables, splitting values, and terminal node (leaf) estimates. As in linear models, the priors allow for regularization, while the Bayesian framework allows for posterior averaging and inference using MCMC. However, like other tree-based methods, BART has a number of advantages over linear models in that it automatically attempts to account for non-linearities and interactions between variables. For binary dependent variables, BART is simply modified by probit transforming the outputs such that they are on the probability scale \cite{chipman2010}.  Through simulations and applications, \cite{chipman2010} showed BART to be highly competitive against advanced machine learning methods in prediction problems. \cite{hill2011} further demonstrated the efficacy of BART when used as an outcome model for causal inference. Though Hill's paper dealt with relatively few variables, \cite{chipman2010} showed that BART scales quite well to high-dimension regression problems. 

A sum-of-trees modeling approach to estimation of propensity scores resembles the generalized boosted modeling (GBM) approach used by \cite{mccaffrey2004} in their application of boosting to propensity score weighting. Like GBM, BART is data adaptive and works well in high-dimensional settings with little user input, but it also provides posterior inference via MCMC \cite{hill2011}. For these reasons, BART is an appealing Bayesian option for estimating a propensity score model. Further details of the algorithm and our implementation are included in the supplementary appendix.

\subsection{Propensity Score Weighting and Competitors}

When used for a treatment regression, the fitted values from regularized Bayesian logistic models and BART are samples from the propensity score distribution. Weighting is a typical way of using propensity scores to get unbiased ATE estimates. Propensity score weighting creates a pseudo-population of subjects from the original sample in which the treatment effect is unconfounded by covariate imbalances between treatment groups. A weighted estimator is given by:  
 
\begin{equation}
{\Delta}_{\tiny \mbox{IPW}} = \bigg(\sum_{i=1}^n \frac{X_i}{{\pi}_i}\bigg)^{-1} \sum_{i=1}^n \frac{X_i Y_i}{{\pi}_i} - \bigg(\sum_{i=1}^n \frac{1-X_i}{1-{\pi}_i}\bigg)^{-1} \sum_{i=1}^n \frac{(1-X_i) Y_i}{1-\pi_i}
 \label{IPW_estimator}
\end{equation}

\noindent where for brevity ${\pi}_i = {\pi}(\mathbf{C}_i)$.

Theoretical results and simulations studies show weighting to be superior to matching or regression adjustment in some settings, though matching can sometimes be more stable and intuitive in practice \cite{austin2010}\cite{lunceford2004}. Weighting is also relatively interpretable and easy to use, garnering an increasingly large user base among health care researchers \cite{austin2015}. For these reasons and to avoid explicit modeling of the outcome regression, we develop below a weighted estimator with close connections to equation (\ref{IPW_estimator}). Augmented inverse probability weighting (AIPW) comprise a doubly-robust and sometimes more efficient modification to equation (\ref{IPW_estimator})  \cite{robins94}. Recent work has created Bayesian techniques for doubly-robust estimation:  \cite{saarela2016} discuss theoretical connections and hurdles to Bayesian doubly-robust estimation, \cite{graham2016} present an approximately Bayesian method based on the Bayesian bootstrap, and \cite{cefalu2016} describe an algorithm for model-averaged doubly-robust estimates.  Targeted maximum likelihood is another doubly-robust alternative, though one that has been researched solely in a frequentist framework \cite{vanderlaan2011}. 

In terms of inference, multiple proposals exist to derive the variance of equation (\ref{IPW_estimator}), including sample variances and the empirical sandwich method \cite{lunceford2004}, as well as the non-parametric bootstrap. Advantages of the bootstrap are that it works with arbitrary propensity score models and can account for the uncertainty associated with the weights. Simulation studies have shown that bootstrapping propensity score weighting is competitive with the sandwich variance and can outperform it in some settings \cite{saarela2015}\cite{bodory2016}. Bootstrapping provides approximate Bayesian posteriors with non-informative priors and, by refitting the treatment regression on bootstrapped data, propagates uncertainty from estimation of the propensity scores \cite{bodory2016}. In this way, bootstrapping serves as a bridge between typical propensity score estimators and methods that use more formal Bayesian models for propensity scores  \cite{saarela2015}\cite{graham2016}.   

\subsection{Bayesian Propensity Score Weighting} \label{sec:bayesian_ps_weighting}

\indent

The estimator in (\ref{IPW_estimator}) generates a weighted pseudo-population by allowing every subject to represent $w_i = \frac{X_i}{\pi_i} + \frac{1-X_i}{1-\pi_i}$ subjects. Intuitively we are taking a difference in means where treated subjects who ``look like" untreated subjects (and vice versa) receive more weight, leading to an unconfounded point estimate of the causal effect \cite{lunceford2004}.  Because $w_i > 1$, the weights are re-normalized such that the size of each treatment group in the pseudo-population is equal to the size of the original treatment group. 


After estimating the propensity score model, we create an unconfounded pseudo-population by weighting each subject using $w_i$. In this pseudo-population, let $\widetilde{Y}(0)$ denote the total number of events in the control group and $\widetilde{Y}(1)$ denote the total number of events in the treatment group, with fixed population sizes $n_0$ and $n_1$ respectively. We use a binomial likelihood $\widetilde{Y}(1) \mid p_1 \sim \mbox{Binomial}(n_1, p_1)$ where $p_1$ is the marginal probability of the event in the treatment group. For the control group we analogously have $\widetilde{Y}(0) \mid p_0 \sim \mbox{Binomial}(n_0, p_0)$. We are interested in inference on the marginal outcome probabilities $p_0$ and $p_1$, which are equal in expectation to the quantities in equation (\ref{IPW_estimator}) and thus unconfounded. So using Bayes' rule and suppressing the fixed $n_0$ and $n_1$ their posteriors are given by: 
\begin{align}
\Prob(p_0 \mid \widetilde{Y}(0)) \propto \Prob(\widetilde{Y}(0) \mid  p_0) \cdot \Prob(p_0)~~\mbox{and}~~
\Prob(p_1 \mid \widetilde{Y}(1)) \propto \Prob(\widetilde{Y}(1) \mid  p_1) \cdot \Prob(p_1).
\end{align}
For the priors, we assume independent beta distributions $p_0 \sim \mbox{Beta}(\alpha_{00}, \alpha_{01})$ and $p_1 \sim \mbox{Beta}(\alpha_{10}, \alpha_{11})$. The hyperparameters, $\{\alpha_{00}, \alpha_{01}, \alpha_{10}, \alpha_{11}\}$, correspond to, respectively, prior counts of no outcomes in the control group, counts of outcomes in the control group, counts of no outcomes in the treatment group, and counts of outcomes in the treatment group. We set $\{\alpha_{00}, \alpha_{01}, \alpha_{10}, \alpha_{11}\} = \{1,1,1,1\}$, implying a flat prior distribution over [0,1] for $p_0$ and $p_1$, though informative hyperparameters could be specified if appropriate. The beta priors are conjugate to the binomial likelihood and closed form posteriors for $p_0$ and $p_1$ are obtained by augmenting the prior counts with those observed in the data. Thus for a given propensity score $\boldsymbol \pi (\mathbf{C})$ and observed data $\mathbf{D} = \{\boldsymbol Y, \boldsymbol X\}$ corresponding to the $n \times 2$ matrix of observed binary outcome and treatment vectors, estimated posterior ${p}_0$ and ${p}_1$ are given by: 
\begin{align}
{p}_1 \mid \mathbf{D}, { \boldsymbol \pi}(\mathbf{C}) \sim \mbox{Beta}  \big(a_1, b_1 \big) ~~\mbox{and}~~
{p}_0 \mid \mathbf{D}, { \boldsymbol \pi}(\mathbf{C}) \sim \mbox{Beta}  \big(a_0, b_0 \big),\label{Y_x}
~ \mbox{where}
\end{align}
\begin{alignat*}{2}
&a_1 = \alpha_{11} +  \gamma_1 \bigg(\sum_{i=1}^n \frac{X_i Y_i}{\hat{\pi}_i}\bigg) ; ~ &&a_0 = \alpha_{00} + \gamma_0 \bigg(\sum_{i=1}^n \frac{(1-X_i) Y_i}{1-\hat{\pi}_i} \bigg);\\
&b_1 = \alpha_{10} + \gamma_1 \bigg(\sum_{i=1}^n  \frac{X_i (1-Y_i)}{\hat{\pi}_i} \bigg);  ~ &&b_0 = \alpha_{01} + \gamma_0 \bigg(\sum_{i=1}^n \frac{(1-X_i) (1-Y_i)}{1-\hat{\pi}_i} \bigg);\\
&\gamma_1 = \frac{\sum_{i=1}^n X_i}{\sum_{i=1}^n X_i / \hat{\pi}_i}; ~ &&\gamma_0 = \frac{\sum_{i=1}^n (1-X_i)}{\sum_{i=1}^n (1-X_i) / (1-\hat{\pi}_i)}.
\end{alignat*}
\noindent
with $\hat{\pi}_i = \hat{\pi}(\boldsymbol{C}_i)$.
The parameter $a_1$ is the prior number of subjects with events ($\alpha_{11}$) plus the number of subjects with events in group 1 of the pseudo-population, and $b_1$ is the prior number of subjects having no events ($\alpha_{10}$) plus the number of subjects with no events in the group 1 pseudo-population. Both are renormalized by $\gamma_1$ so that their sum is equal to the size of the original (unweighted) group 1 population. Similar interpretations hold for $a_0$, $b_0$, and $\gamma_0$ in group 0. With the specification $\{\alpha_{00}, \alpha_{01}, \alpha_{10}, \alpha_{11}\} = \{1,1,1,1\}$ the priors add very little information. 

We can draw $J$ times from the outcome model for each of $K$ draws from the propensity score model. The $jth$ draw from the estimated posterior distribution of the causal effect, ${\Delta}$, using the $kth$ draw for $ \hat{{\boldsymbol \pi}}^k$ can be calculated as:
\begin{equation}
\hat{{\Delta}}^{jk} \mid \mathbf{D}, \hat{{\boldsymbol \pi}}^k (\mathbf{C})= \hat{p}^{jk}_1 - \hat{p}^{jk}_0. \label{delta}
\end{equation}
Because (\ref{Y_x}) is conditional on a single value of $\hat{\boldsymbol \pi}^k (\mathbf{C})$, we can evaluate ${\Delta}^{jk} | \mathbf{D}, \hat{\boldsymbol \pi}^k (\mathbf{C})$ over the distribution of $\boldsymbol \pi (\mathbf{C})$. \cite{kaplan2012} proposed combining the estimates using the law of total variance to get: 
\begin{align}
\mathbb{E}({\Delta} \mid \mathbf{D})  &= \mathbb{E}_{\pi}\{\mathbb{E}({\Delta} \mid \mathbf{D}, \boldsymbol \pi)\} ~\mbox{and}~
\mathbb{V}({\Delta} \mid \mathbf{D})  = \mathbb{E}_{\pi}\{\mathbb{V}({\Delta} \mid \mathbf{D}, \boldsymbol \pi)\} + \mathbb{V}_{\pi}\{\mathbb{E}({\Delta} \mid \mathbf{D}, \boldsymbol \pi)\}, \label{total_variance}
\end{align}
which resembles multiple imputation. Formula (\ref{total_variance}) demonstrates how the variance from each model contributes to the final uncertainty: the average variance from the outcome models enters as the first term in (\ref{total_variance}) and variance across the propensity scores enters as the second term. With our proposed algorithm we get draws from the outcome model for each propensity score draw, not just a variance estimate. Thus we recommend generating the entire posterior density ${\Delta} \mid \mathbf{D}$ by simply concatenating across propensity score draws. The variance can then be computed directly from these draws and uncertainty measures like 95\% confidence intervals can be calculated from their empirical quantiles.

\cite{saarela2015} critiqued the approach of \cite{kaplan2012} for using a non-likelihood outcome model when generating the posterior variance in equation (\ref{total_variance}). Because we specify a weighted binomial likelihood for the counts $\widetilde{Y}(0)$ and $\widetilde{Y}(1)$, our approach is grounded in the theoretical work of \cite{saarela2015}, who used propensity score weights along with a Dirichlet distribution for Bayesian bootstrapping to obtain posterior draws. However in contrast to the method of \cite{saarela2015}, our approach takes advantage of conjugacy and has clear connections to more conventional IPW.       

In our supplementary materials, we show the equivalence between the posterior mean of equation (\ref{delta}) and the IPW estimator in equation (\ref{IPW_estimator}). We also provide further computational details and R code for sampling from $\Delta \mid \mathbf{D}$.

\section{Simulation Studies}
\label{sec:simulations}

\subsection{Handling of Propensity Score Distribution}

\indent

\cite{saarela2015} made the case that even in a Bayesian framework, propensity scores should be fixed to their maximum likelihood estimates or posterior mean estimates rather than integrating over the full distribution. However, doing so in high-dimensions ignores the crucial source of uncertainty from estimating the propensity scores, which could potentially cause under coverage and result in more type I errors. To test this, we first constructed some very simple simulation scenarios. We let $\boldsymbol X \sim \mbox{Bernoulli}\big\{\boldsymbol (\mbox{logit}^{-1} (.5 \boldsymbol C_1 + .5 \boldsymbol C_2) \big\}$ and $\boldsymbol Y \sim \mbox{Bernoulli}\big\{ \mbox{logit}^{-1} (\boldsymbol X - .5 \boldsymbol C_1 - .5 \boldsymbol C_2) \big\}$ where $\boldsymbol C_1$ and $\boldsymbol C_2$ were drawn from independent standard normals with $n = 100$ subjects. We specified three Bayesian logistic models to estimate the propensity scores. The first estimates the correct model, with $\boldsymbol C_1$ and $\boldsymbol C_2$ as confounders. The second estimates an ``over specified" model that includes $\boldsymbol C_1$, $\boldsymbol C_2$, and 8 additional covariates with no relation to treatment or outcome. Finally, we estimate an ``under specified" model that includes only $\boldsymbol C_1$, making $\boldsymbol C_2$ an unmeasured confounder. The results after 4000 simulations appear in table \ref{small_simulation_table}.

In their original paper, \cite{saarela2015} focused on analyzing propensity score estimators across across different levels of confounding. They found that including variance from the propensity score estimate tended to cause slight over coverage in all scenarios, while IPW and fixed propensity score Bayesian estimators had either perfect coverage or  under covered the true parameter.  We too find that integrating over the propensity score distribution leads to slight over coverage, making it slightly conservative. Fixing the propensity score parameter estimates to the posterior means leads to good coverage when there are few variables, but underestimates the variance in the over specified case and is slightly more biased. Because we are primarily concerned with over specified models in high-dimensions, in which there are likely to be many noisy variables, the robust coverage of the integrated method in this setting is key. Thus these initial simulations supported integrating over the propensity score distribution for high-dimensional analysis.  

\begin{table}[!h]
	\centering
	\caption{\footnotesize{\textit{Properties of Bayesian propensity score weighting over 4000 simulations where propensity scores are either integrated over their posterior distribution or fixed to their posterior means. Sample size within each simulation is $n=100$. Coverage of 95\% posterior intervals is reported.}}}
	\begin{tabular}{llrr}
		& & Integrated PS & Mean PS \\ 
		\hline
		& Bias & .0008 & -.0069 \\
		Correct &	Variance & .0102 & .0086 \\
		& Coverage & 96.5\% & 94.7\% \\
		\hline
		& & Integrated PS & Mean PS \\ 
		\hline
		& Bias & .0006 & -.0270 \\			
		Overspecified &	Variance & .0157 & .0085 \\
		& Coverage & 96.6\% & 91.1\% \\ 
		\hline
		& & Integrated PS & Mean PS \\ 
		\hline
		& Bias & -.0519 & -.0566 \\			
		Underspecified &	Variance & .0094 & .0087 \\
		& Coverage & 91.9\% & 90.4\% \\ 
		\hline
	\end{tabular}
	\label{small_simulation_table}
\end{table}

\subsection{More Realistic Simulations}
\indent

We compare three Bayesian methods to estimate the propensity score distribution: a logistic regression with weakly informative priors defined as $\mbox{Student-t}_3(0, 2.5)$ \cite{gelman2014}, a logistic regression regularized by horseshoe priors, and BART with $m = 200$ trees grown using default priors on the tree depth and end nodes (details in appendix). We altered the horseshoe method slightly by replacing the half-Cauchy priors with half-$t_3(0,1)$ priors, which allow much easier fitting using Monte Carlo sampling without substantive difference in estimates \cite{stan}. A naive estimate is computed as the unadjusted difference in means between treatment groups. As a baseline competitor, we examine IPW as in equation (\ref{IPW_estimator}) with empirical sandwich standard errors. We also compare to targeted maximum likelihood estimation (TMLE), a frequentist approach to causal inference in high-dimensional settings \cite{vanderlaan2011}. TMLE used the SuperLearner ensemble method with unregularized and lasso penalized logistic regression as cross-validated candidate models to fit outcome and propensity score regressions. These regressions are combined to formulate a doubly-robust estimate of the ATE \cite{vanderlaan2011}. 

We ran 500 simulations with $n = 1000$ and $p = 100$ constructed to partly resemble our coronary stent data in terms of coefficient values and confounder distributions, while remaining computationally tractable. To this end, 100 binary confounders were extracted from the data, fit to the treatment variable using logistic regression, and rounded to the nearest 10th. These estimated coefficients were then used as the true treatment coefficients ($\boldsymbol{\beta}^X$) in simulated datasets. The values of $\boldsymbol{\beta}^X$ ranged from -1.1 to 1.1 on the log-odds scale, with a lower quartile of -0.2 and upper quartile of 0.2, making most of the coefficients quite small, though only 18 were exactly equal to 0. The confounders $\mathbf{C}$ were drawn from independent Bernoulli distributions with probabilities equal to their empirical prevalence in our data. As a result, many of the covariates were quite sparse. The outcome coefficients $\boldsymbol{\beta}^Y$ were then selected to create confounding while enforcing $\sum \boldsymbol{\beta}^Y = 0$. The data were simulated following: $\Prob(X_i | \boldsymbol C_i) \sim \mbox{Bernoulli} \{ \mbox{logit}^{-1} (\beta_0^X + \sum_{j = 1}^p \beta^X_j C_{ij})  \}$, yielding a marginal treatment probability of $\Prob(X_i = 1) \approx .7$, and $\Prob(Y_i |  X_i, \boldsymbol C_i) \sim \mbox{Bernoulli} \{ \mbox{logit}^{-1} (\beta_0^Y + \beta^{\footnotesize\mbox{Tr}} X_i + \sum_{j = 1}^p \beta^Y_j C_{ij}) \}$. We set $\beta^{\footnotesize\mbox{Tr}} = -2$, yielding an ATE over $\mathbf C$ of $\Delta \approx -0.15$. An outcome intercept of $\beta_0^Y = -2$ gave a relatively low event rate with $\Prob(Y_i = 1) \approx 0.1$. We assessed bias, mean squared error (MSE), confidence interval width, and realized 95\% posterior interval coverage of each method. 

Weakly informative $t_3(0,2.5)$ priors had almost perfect coverage, but the highest MSE of any method (Table \ref{realistic_simulation_table}). The lack of regularization likely made the treatment regressions unstable, and IPW suffered from similarly poor MSE results. The horseshoe and BART faired better, with considerably improved MSE, tight confidence intervals, and reasonably good coverage. BART even improved slightly over the horseshoe in terms of bias and coverage. It's competitive performance is remarkable because, unlike the logistic regression-based approaches which correctly specified the model as linear, BART is completely non-parametric. 
 
Fixing the propensity scores to their posterior means considerably reduced variance estimates, as evidenced by tighter confidence intervals for all methods, but led to poor coverage as a result. Although the MSE of the $t_3(0,2.5)$ priors improved, the MSE and bias were actually slightly higher for the other Bayesian methods. As expected, the naive difference in means performed very poorly. The IPW suffered from poor MSE and slightly under-covered the true treatment effect. While the TMLE had among the best performance with the lowest MSE, low bias, and tight confidence intervals, it under-covered the true treatment effect considerably, indicating the inadequacy of it's standard error estimate. Bootstrapping TMLE would likely yield better coverage by incorporating uncertainty from estimating the scores, but was computationally infeasible in this simulation setting.

Our simulations show the effectiveness of Bayesian propensity scores and support including uncertainty from the propensity score distribution when fitting large treatment models. We also demonstrated the utility of Bayesian methods for regularization and non-parameteric regression for reducing the MSE of high-dimensional causal estimates.

\begin{table}[!h]
	\centering
	\caption{\footnotesize{\textit{Bias (estimate - true effect), mean squared error times 1000 (MSE $\times 10^3$), average confidence interval (CI) width, and coverage of 95\% posterior intervals for various estimators over 500 simulations with sample size $n=1000$ and $p=100$ confounders specified above. Bayesian propensity scores were implemented with different treatment of the propensity score distribution (integrated or mean), and different modeling choices (horseshoe priors, $t_3(0,2.5)$, or Bayesian additive regression trees (BART)). A naive difference in means, classic inverse probability weighting with robust standard errors (IPW), and targeted maximum likelihood (TMLE) were also evaluated.}}}
	\begin{tabular}{lrrrr}
		
		Treatment Model & Bias & MSE $\times 10^3$ & CI Width & Coverage \\ 
		\hline
		\textit{Integrated Propensity Score} & & & &  \\
		\hline
		Logistic, $t_3(0,2.5)$ Priors& -.011 & 2.82 & .220 & 95.2\% \\
		Logistic, Horseshoe Priors& .016 & 0.93 & .110 & 93.0\% \\
		BART, Default Regularizing Priors & .011 & 0.84 & .123 & 96.8\% \\
		\hline
		\textit{Mean Propensity Score} & & & &  \\
		\hline
		Logistic, $t_3(0,2.5)$ Priors& -.001 & 1.59 & .095 & 79.2\% \\
		Logistic, Horseshoe Priors& .018  & 0.96 & .092 & 86.0\% \\
		BART, Default Regularizing Priors & .015 & 0.86 & .093 & 87.2\% \\
		\hline
		\textit{Other Methods} & & & & \\
		\hline
		IPW & -.001 & 1.67 & .151 & 92.8\% \\
		TMLE & .006 & 0.78 & .075 & 81.4\% \\ \hline
		Naive Estimate & .030 & 1.46 & .092 & 73.0\% \\
		\hline
	\end{tabular}
	\label{realistic_simulation_table}
\end{table}
  
\section{Application: Drug Eluting vs Bare Metal Coronary Stenting in Massachusetts}
\label{sec:mass_dac}
\indent

We revisit a comparison of the causal effect of using drug-eluting stents (DES) to bare-metal stents (BMS) on the risks of all-cause mortality and a composite endpoint including mortality, acute myocardial infarction, and target vessel revascularization (TVR) within 1 year after the initial procedure. Past studies using propensity score techniques have consistently found a benefit to drug-eluting stents on lowering rates of 1-year mortality.  In contrast RCTs and an instrumental variables analysis report no such benefit \cite{mauri_circ2008} \cite{venkitachalam2011} \cite{bonaa2016}. However, these analyses have utilized well fewer than 100 variables for confounding adjustment. An analysis using a richer confounder set may align more closely with experimental results. Although we are primarily interested in TVR, we use a composite endpoint, which has been used in past stent trials, to avoid bias due to competing risks.


We examined $n = 8718$ patients from the Mass-DAC clinical registry consisting of percutaneous coronary interventions (PCIs) performed in Massachusetts adults between October 1, 2010 and September 30, 2011 inclusive. We merged the Mass-DAC data with hospital billing data from the Massachusetts Center for Health Information and Analysis (CHIA). The Mass-DAC data contributed 131 potential confounders, a mix of measures indicating patient characteristics (e.g. age, sex), pre-existing conditions (e.g. diabetes, heart failure), or characteristics of the procedure (e.g. treated vessel, hospital). There were 12 continuous covariates, the remainder being binary or categorical covariates. The CHIA data included 15 free-response fields containing Present On Admission (POA) diagnoses entered as International Classification of Diseases, 9th Version (ICD-9) codes. POA codes describe patients at hospital arrival and therefore not impacted by treatment decisions. To set a lower bound on sparsity we considered all diagnoses for which 10 or more patients were coded as having the condition and converted them to dummy variables.  This yielded 364 additional covariates for a total of 495 potential confounders. Henceforth we refer to the combined dataset as Mass-DAC. We imputed a small amount of missing data, creating one imputed dataset for simplicity. Due to the small amount of missingness substantive inferences would be unlikely to change if we had used multiple imputed datasets. 
 
We estimate the ATE using Bayesian propensity score weighting with treatment models estimated from the full Mass-DAC data. Our causal parameter is defined as the average event rate in the BMS group subtracted from the average event rate in the DES group. We focus on analyzing two outcomes: 1-year mortality and a composite endpoint of 1-year mortality, acute myocardial infarction (AMI), and TVR. We derived 1-year AMI from records of in-hospital events as mandated in the Mass-DAC registry, or from any within-year hospital readmission with diagnosis code 410.x1 in the CHIA data. We defined 1-year TVR as a PCI on the same vessel or any coronary artery bypass graft (CABG) surgery within 1 year of stent implantation. As an additional check, we include 30-day mortality as a falsifiability endpoint. Since a true causal effect on mortality is unlikely to manifest after 30-days, we know a significant difference on this outcome is attributable to residual confounding. 
 
For the first stage we estimated Bayesian models with three different specifications: Student-t priors, horseshoe priors, and BART with 2000 trees grown using default parameters. The logistic regressions were fit using Stan and convergence was assessed by inspecting trace plots and by ensuring all within-chain relative to between-chain statistics (R-hat) were less than 1.1 \cite{stan}. BART was fit in R using the package BayesTree, written by the original authors \cite{chipman2010}. We implemented an IPW estimator as in equation (\ref{IPW_estimator}) where propensity scores were estimated with logistic regression and the variance estimate used the empirical sandwich method or 2000 bootstrap replicates \cite{austin2015}. As a final competitor, we implemented TMLE with SuperLearner using unregularized and lasso penalized logistic regression as candidate models, with variance estimated from the influence curve or from 1000 bootstrap replicates \cite{vanderlaan2011}. We denote the sandwich variance IPW and influence curve variance TMLE  with the preface {\em analytic}, to reflect the closed-form nature of their variance estimates. We expect the analytic methods to have tighter confidence intervals because they do not reflect variance from the propensity score model.

We assess positivity, balance, and stability of our estimated propensity scores. We also examine the distributions of weights and the (weighted) absolute standardized differences in covariates between the DES and BMS groups \cite{austin2015}. High weights tend to result from an overfit propensity score model and can also indicate a positivity violation, though low weights are no guarantee of good overlap \cite{petersen2010}. The standardized difference in covariates provides a measure of balance, where an absolute standardized difference of less than 10\% is considered negligible \cite{austin2015}. For the Bayesian methods, we graphically examined the distributions of balanced standardized differences in order to assess post-weighting covariate balance as well as the stability of the estimates. Optimally, the 95\% posterior intervals for the standardized difference would be within $(-10,10)$ for all confounders, though in analogy to standard propensity score analyses balance is acceptable if all posterior means are within $(-10,10)$ \cite{kaplan2012}. Finally, we assessed the distribution of propensity score weights across subjects for both our Bayesian and non-Bayesian methods, defined as $X_i/\pi_i + (1-X_i)/(1-\pi_i) $. For the Bayesian methods the analyzed weights correspond to each subjects posterior mean weight, while for the frequentist methods there is only 1 weight (from the maximum likelihood propensity score) returned for each person.

\subsection{Results}

All Bayesian methods converged satisfactorily according to the trace plots and R-hat statistics. All TMLE bootstrap iterations fit, though about 25\% of the bootstrapped IPW estimates failed to converge as a result of the data sparsity and lack of regularization.  

\begin{figure}[!h]
	\begin{center}
		\includegraphics[trim = 0mm 0mm 0mm 0mm, clip, height=3.8in]{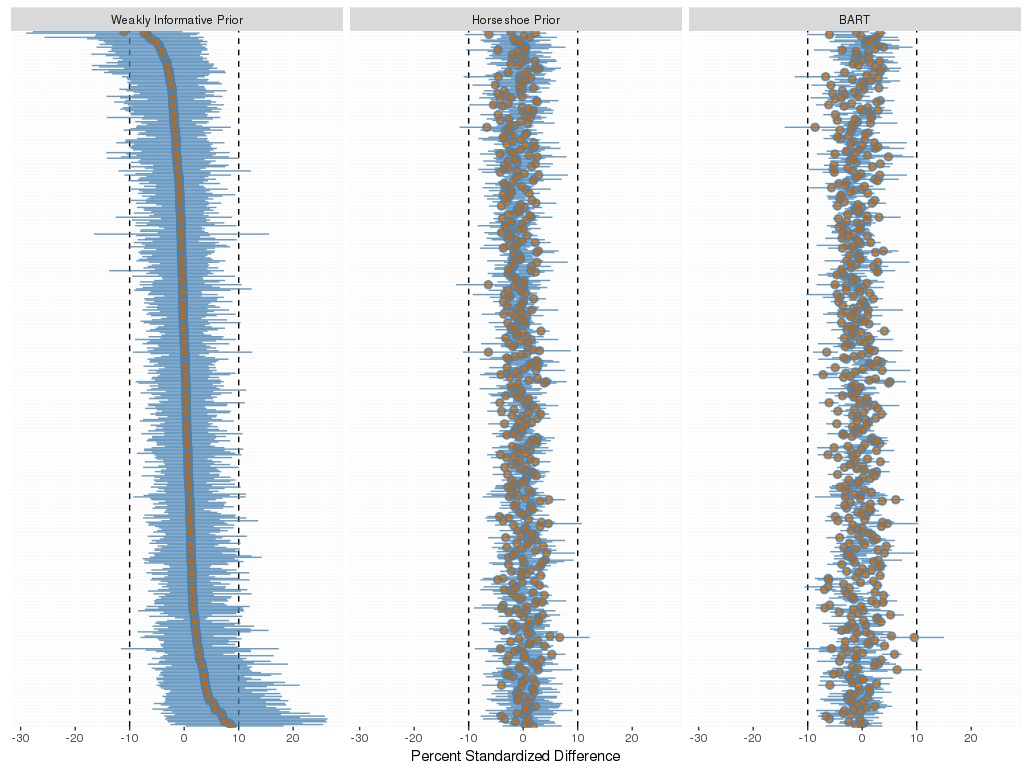}
		\caption{\footnotesize{\textit{Mean (orange) and 95\% posterior interval (blue) of standardized difference for each confounder after balancing on each draw from the propensity score model. Each panel represents a different propensity score model specification. The dotted lines at -10 and 10 indicate range usually taken to be acceptable. Points are sorted by the mean standardized difference from the weakly informative $t_3(0,2.5)$ specification.}}}
		\label{fig:std_diff}
	\end{center}
\end{figure}

\indent
Figure \ref{fig:std_diff} shows the distribution of the percent standardized differences for each of the 495 confounders for the Bayesian propensity score methods, sorted by the mean under weakly informative $t_3(0,2.5)$ priors. The $t_3(0,2.5)$ priors for the logistic treatment regression coefficients show a significant portion of the standardized difference distributions for some confounders outside of the -10\% to 10\% range. In contrast, the standardized difference distributions under horseshoe priors for the logistic treatment regression coefficients and those using a BART treatment model were almost all entirely within an acceptable range. Due to their heavier regularization, the posteriors under horseshoe priors and BART are more stable. The fact that some parts of the posterior under $t_3(0,2.5)$ priors do not sufficiently balance the confounders is concerning. The $t_3(0,2.5)$ priors had the highest average and maximum weights of all the approaches, while horseshoe priors and BART had the lowest. None of the methods' weights indicated a positivity violation.

\begin{sidewaystable}[!p]
	\centering
\caption{\footnotesize{\textit{Mean (max) percent absolute standardized difference (Std Diff) across covariates, mean (max) propensity score weights, estimates of causal effects (95\% confidence interval) as percent difference in outcome rate between DES and BMS, $\Delta = p_1 - p_0$, after adjustment. A negative estimate indicates DES yields a lower rate than BMS. IPW = inverse probability weighting; TMLE = targeted maximum likelihood.}}}
	\begin{tabular}{lrrrr}
		\hline \\ 
		                    &                & \multicolumn{2}{c}{\textbf Mortality} & \\ \cline{3-4}
		 \textbf{Approach} & \textbf{Weight} & \textbf{30-day} & {\textbf 1-Year} & \textbf{1-Year Composite }\\
		 \hline
	 {Naive} & NA & -3.4 (-4.1, -2.6) & -6.9 (-8.0, -5.7) & -9.4 (-11.1, -7.6) \\ \hline
	 Analytic TMLE & NA & -1.1 (-1.7, -0.6) & -3.0 (-4.0, -2.0) & -7.3 (-9.0, -5.5) \\
	 ~~{Bootstrapped TMLE}	& NA & -1.1 (-2.7, -0.5) & -3.0 (-4.4, -1.8) & -7.3 (-10.0, -5.2) \\
	 {Analytic IPW} & 2.0 (54.9) & -1.0 (-2.0, -0.1) & -2.9 (-4.3, -1.5) & -6.9 (-9.3, -4.4) \\
	 ~~{Bootstrapped IPW}	& 2.0 (54.9) & -1.0 (-2.0, 1.8) & -2.9 (-4.9, -0.3) & -6.9 (-10.1, -3.1) \\ 
	 \hline
	 Logistic, $t_3(0,2.5) Priors$ & 2.3 (100.6) & -0.3 (-1.7, 2.5) & -2.1 (-4.2, 0.9) & -6.1 (-9.3, -2.4) \\
	 Logistic, Horseshoe Priors & 2.0 (29.6) & -1.2 (-2.0, -0.5) & -3.2 (-4.5, -2.0) & -7.6 (-9.6, -5.5)  \\
	 {BART}, Default Regularizing Priors& 2.0 (28.2) & -1.3 (-2.1, -0.6) & -3.3 (-4.5, -2.1) & -7.4 (-9.3, -5.5) \\
	 \hline
	\end{tabular}	
	\label{table:mdac_estimates}
\end{sidewaystable}

Table \ref{table:mdac_estimates} displays the point estimates and 95\% posterior intervals for each method and outcome. In terms of the estimated causal effect, every method sharply decreased the magnitude of the naive 1-year (30-day) mortality difference from 6.9 (3.4) while also decreasing the magnitude of the composite endpoint difference from 9.4. Even after adjustment, most methods found a significant benefit to mortality for DES. This is particularly concerning for 30-day mortality as any differences observed that soon after implantation are almost certainly a result of residual confounding.  Only bootstrapped IPW and Bayesian propensity scores with $t_3(0,2.5)$ priors found no significant DES benefit to 30-day mortality, though the other methods found only a small benefit. Only the Bayesian logistic treatment model with $t_3(0,2.5)$ priors found no benefit to 1-year mortality. These findings may reflect residual confounding, as past RCTs and instrumental variables analyses have repeatedly confirmed no mortality benefit \cite{venkitachalam2011} \cite{bonaa2016}. However, our methods did adjust away considerably more confounding than \cite{mauri_circ2008}, which also used the Mass-DAC data. As expected, all methods found a significant DES benefit to the composite endpoint, as shown in clinical trials.

\begin{figure}[!h]
	\begin{center}
		\includegraphics[trim = 0mm 0mm 10mm 10mm, clip, height=3.8in]{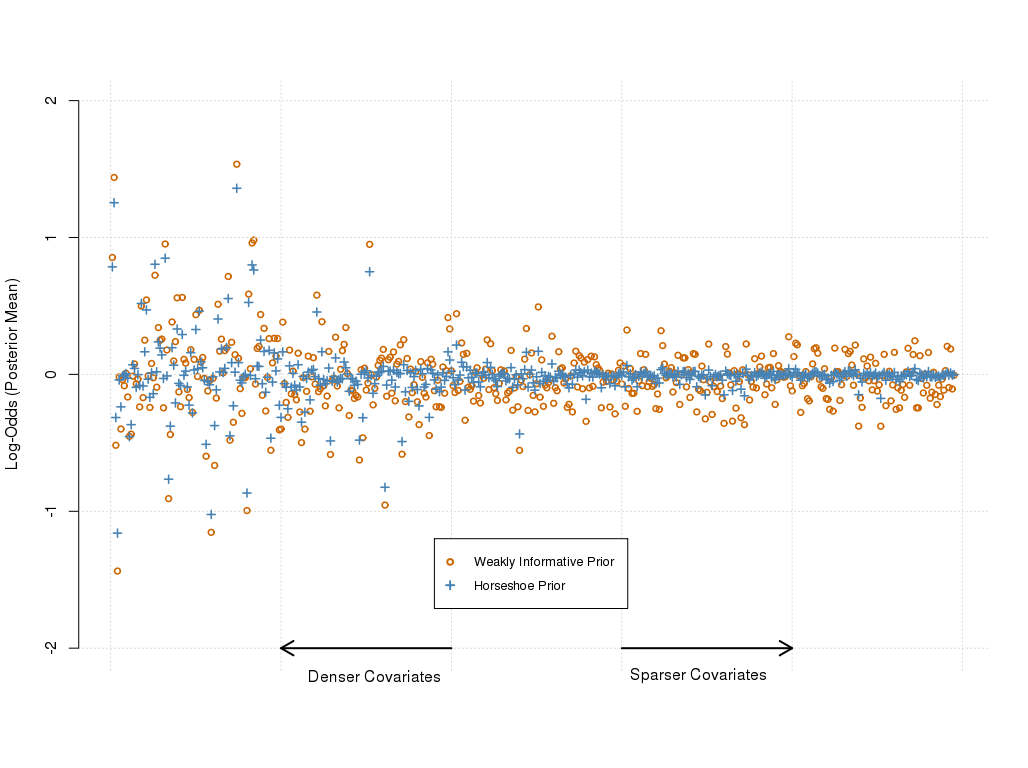}
		\caption{\footnotesize{\textit{Posterior mean coefficients on log-odds scale, estimated from the Mass-DAC propensity score model for weakly informative $t_3(0,2.5)$ and horseshoe priors. The 12 continuous covariates appear on the far left, and the remaining binary indicators are sorted in order of decreasing density. Thus indicators observed in many subjects appear towards the left side of the plot and rarely observed indicators appear towards the right. Continuous confounders were standardized before fitting for compatibility with regularization priors, binary coefficients were standardized only for this display.}}}
		\label{fig:mdac_coefs}
	\end{center}
\end{figure}

Figure \ref{fig:mdac_coefs} compares standardized posterior mean coefficient estimates for the Bayesian logistic regression treatment model with $t_3(0,2.5)$ priors and horseshoe priors. Sparser binary covariates appear to the right and are shrunk significantly towards zero by the horseshoe priors. The horseshoe priors may be shrinking the coefficients of sparse binary covariates too aggressively, as they may be rare but important confounders. On the other hand, imposing little regularization can be problematic as well, evidenced by our simulation studies, the instability apparent in figure \ref{fig:std_diff}, and the fact that almost 500 of our 2000 bootstrapped IPW estimates failed to converge. 

BART provided covariate balance and estimates of the risk difference comparable to the horseshoe priors. In this example, accommodating non-linearities and interactions did not make a large difference on the causal estimate compared to the main effects logistic models.

\section{Discussion}

\indent

The methods we discussed combine the advantages of a Bayesian framework and frequentist propensity scores for causal inference with many variables, where including as many confounders as possible will help meet the ignorability assumption. In particular, Bayesian logistic regression using Student-$t$ priors or horseshoe priors and BART provide useful parametric and non-parametric options to estimating propensity score distributions from high-dimensional data sets. Through simulations we showed that integrating over the propensity score model can slightly reduce bias and mean squared error but significantly improve coverage in high-dimensional settings. Because  a significant portion of the variance in a propensity score analysis stems from estimating the scores, incorporating uncertainty from the propensity score regression was important to achieving good coverage in our Bayesian approaches. We also found that IPW coupled with bootstrapping gave more reasonable inferential results in our application than the empirical sandwich method, which does not account for uncertainty in the propensity score model.

We proposed horseshoe priors and BART as techniques for variance reduction and flexible model fitting. While these approaches dominated more conventional Student-$t$ priors in our simulations and looked better diagnostically in our application, they failed to meet our falsifiability endpoint. For causal inference, we recommend weakly informative priors for known confounders, airing on the side of reducing bias over variance, incorporating substantive knowledge for variable selection and prior specification when available, and using sparsity-inducing priors when there is noisy data. A Bayesian framework provides a comprehensive and intuitive way to build models that include substantive knowledge and regularization, as well as adaptive specifications when little is known about the data generating process.

In our application we assessed the benefits of DES over BMS on preventing mortality and a composite endpoint. We found that by including a rich set of almost 500 confounders for adjustment and using Bayesian propensity scores we were able to adjust for considerably more confounding than an earlier propensity score analysis with the same data \cite{mauri_circ2008}. When used with weakly informative priors our method aligned with estimates observed in RCTs \cite{bonaa2016}. 

Given that a few approaches exist for causal inference in high-dimensional data settings we have several recommendations. First, although the treatment model is essentially a nuisance parameter, accounting for the uncertainty of who receives what treatment is important in any approach. Second, an outcome free design in which the  treatment model is separated from the outcome model is often desirable and sometimes mandated. Third, making fewer parametric assumptions in the outcome model, for example by using a weighting procedure instead of regression adjustment, is advantageous. Our approach meets these criteria. The methods we discussed did not use any information from the outcome in setting up the design (treatment) model, allowing the user to focus on designing a good propensity score model without revealing the causal effect until balance has been achieved \cite{rubin2008}. Such an approach is not without drawbacks, as including outcome information can aid variable selection or enable doubly-robust estimation \cite{schneeweiss2009}\cite{wang2012}\cite{robins94}\cite{cefalu2016}. Nevertheless, propensity scores remain a powerful and popular tool for causal inference and we believe that our proposed methods provide a useful framework for comparative effectiveness research with high-dimensional data.

\section*{Acknowledgments}

Mr. Spertus' and Dr. Normand's efforts were partially supported by a grant from the National Institute of General Medical Sciences R01-GM111339.  Dr. Normand's effort was also partially supported by a grant from the Center for Devices and Radiological Health U01-FD004493. The authors are indebted to Ann Lovett, R.N., M.A. and Matt Cioffi, M.S., both from Harvard Medical School for preliminary dataset construction and review, and to the Massachusetts Department of Public Health and CHIA for the use of their data. 

\bibliography{CERinBigdataWorksCited}

\begin{thebibliography}{10}

\bibitem{austin2010}
Peter~C Austin.
\newblock The performance of different propensity-score methods for estimating
  differences in proportions (risk differences or absolute risk reductions) in
  observational studies.
\newblock {\em Statistics in Medicine}, pages 2137--2148, 2010.

\bibitem{austin2015}
Peter~C Austin and Elizabeth~A Stuart.
\newblock Moving towards best practice when using inverse probability of
  treatment weighting (iptw) using the propensity score to estimate causal
  treatment effects in observational studies.
\newblock {\em Statistics in Medicine}, 34:3661--3679, 2015.

\bibitem{bodory2016}
Hugo Bodory, Lorenzo Camponovo, Martin Huber, and Michael Lechner.
\newblock The finite sample performance of inference methods for propensity
  score matching and weighting estimators.
\newblock {\em Institute for the Study of Labor: Discussion Paper}, 2016.

\bibitem{bonaa2016}
Kaare~H. Bonaa, Jan Mannsverk, Rune Wiseth, Lars Aaberge, Yngvar Myreng, Ottar
  Nygard, Dennis~W. Nilsen, Nils-Einar Klow, Michael Uchto, Thor Trovik, Bjorn
  Bendz, Sindre Stavnes, Reidar Bjornerheim, Alf-Inge Larsen, Morten Slette,
  Terje Steigen, Ole~J. Jakobsen, Oyvind Bleie, Eigil Fossum, Tove~A. Hanssen,
  Oystein Dahl-Eriksen, Inger Njolstad, Knut Rasmussen, Tom Wilsgaard, and
  Jan~E. Nordrehaug.
\newblock Drug-eluting or bare-metal stents for coronary artery disease.
\newblock {\em The New England Journal of Medicine}, 2016.

\bibitem{stanjstatsoft}
Bob Carpenter, Andrew Gelman, Matt Hoffman, Daniel Lee, Ben Goodrich, Michael
  Betancourt, Michael~A Brubaker, Jiqiang Guo, Peter Li, and Allen Riddell.
\newblock Stan: a probabilistic programming language.
\newblock {\em Journal of Statistical Software}, In Press, 2016.

\bibitem{carvahlo2010}
Carlos~M Carvahlo, Nicholas~G Polson, and James~G Scott.
\newblock The horseshoe estimator for sparse signals.
\newblock {\em Biometrika}, 97:465--480, 2010.

\bibitem{cefalu2016}
Matthew Cefalu, Francesca Dominici, Nils Arvold, and Giovanni Parmigiani.
\newblock Model averaged double robust estimation.
\newblock {\em In submission}, 2016.

\bibitem{bart}
Hugh Chipman and Robert McCullough.
\newblock Package 'bayestree'.
\newblock 2016.

\bibitem{chipman2010}
Hugh~A Chipman, Edward~I George, and Robert~E McCullough.
\newblock Bart: Bayesian additive regression trees.
\newblock {\em The Annals of Applied Statistics}, 4:266--298, 2010.

\bibitem{graham2016}
Emma J~McCoy Daniel J~Graham and David~A Stephens.
\newblock Approximate bayesian inference for doubly robust estimation.
\newblock {\em Bayesian Analysis}, 11:47--69, 2016.

\bibitem{farrell2015}
Max~H Farrell.
\newblock Robust inference on average treatment effects with possibly more
  covariates than observations.
\newblock {\em Journal of Econometrics}, 189:1--23, 2015.

\bibitem{fda2013}
U.S. Food and Drug Administration.
\newblock Best practices for conducting and reporting pharmacoepidemiologic
  safety studies using electronic health records.
\newblock http://www.fda.gov/downloads/drugs/guidances/ucm243537.pdf, 2013.

\bibitem{gelman2014}
Andrew Gelman, John~B Carlin, Hal~S Stern, David~B Dunson, Aki Vehtari, and
  Donald~B Rubin.
\newblock {\em Bayesian Data Analysis: Third Edition}.
\newblock CRC Presss, 2014.

\bibitem{ghosh2015}
Debashish Ghosh, Yeying Zhu, and Donna~L Coffman.
\newblock Penalized regression procedures for variable selection in the
  potential outcomes framework.
\newblock {\em Statistics in Medicine}, 34:1645--1658, 2015.

\bibitem{rstan}
Jiqiang Guo, Jonah Gabry, Ben Goodrich, Daniel Lee, Krzysztof Sakrejda,
  Trustees of~Columbia~University, Oleg Skylar, The R~Core Team, Jens
  Oehlschlaegel-Akiyoshi, Hadley Wickham, Joel~De Guzman, John Fletcher, Thomas
  Heller, and Eric Niebler.
\newblock Package 'rstan'.
\newblock 2016.

\bibitem{hahn2016}
P~Richard Hahn, Carlos~M Carvalho, and David Puelz.
\newblock Regularization and confounding in linear regression for treatment
  effect estimation.
\newblock {\em Bayesian Analysis}, 2016.

\bibitem{hill2011}
Jennifer~L. Hill.
\newblock Bayesian nonparametric modeling for causal inference.
\newblock {\em Journal of Computational and Graphical Statistics}, 20:217--240,
  2011.

\bibitem{kaplan2012}
David Kaplan and Jianshen Chen.
\newblock A two-step bayesian approach for propensity score analysis:
  simulations and case study.
\newblock {\em Pyschometrika}, 77:581--609, 2012.

\bibitem{lunceford2004}
Jared~K. Lunceford and Marie Davidian.
\newblock Stratification and weighting via the propensity score in estimation
  of causal treatment effects: a comparative study.
\newblock {\em Statistics in Medicine}, 23:2937--2960, 2004.

\bibitem{mauri_circ2008}
Laura Mauri, Treacy~S Silbaugh, Robert~E Wolf, Katya Zelevinsky, Ann Lovett,
  Zheng Zhou, Frederic~S Resnic, and Sharon-Lise~T Normand.
\newblock Long-term clinical outcomes after drug-eluting and bare-metal
  stenting in massachusetts.
\newblock {\em Circulation}, 118:1817--1827, 2008.

\bibitem{mccaffrey2004}
Daniel~F McCaffrey, Greg Ridgeway, and Andrew~R Morral.
\newblock Propensity score estimation with boosted regression for evaluating
  causal effects in observational studies.
\newblock {\em Psychological Methods}, pages 403--425, 2004.

\bibitem{mccandless2009}
Lawrence~C McCandless, Paul Gustafson, and Peter~C Austin.
\newblock Bayesian propensity score analysis for observational data.
\newblock {\em Statistics in Medicine}, 28:94--112, 2009.

\bibitem{petersen2010}
Maya~L Petersen, Kristin~E Porter, Susan Gruber, Yue Wang, and Mark~J van~der
  Laan.
\newblock Diagnosing and responding to violations in the positivity assumption.
\newblock {\em Statistical Methods in Medical Research}, 21:31--54, 2010.

\bibitem{robins94}
James~M Robins, Andrea Rotnitzky, and Lue~Ping Zhao.
\newblock Estimation of regression coefficients when some regressors are not
  always observed.
\newblock {\em Journal of the American Statistical Association}, 89:846--866,
  1994.

\bibitem{rosenbaum1983}
Paul~R Rosenbaum and Donald~B Rubin.
\newblock The central role of the propensity score in observational studies for
  causal effects.
\newblock {\em Biometrika}, 70:41--55, 1983.

\bibitem{rubin2008}
Donald~B Rubin.
\newblock For objective causal inference, design trumps analysis.
\newblock {\em The Annals of Applied Statistics}, 2:808--840, 2008.

\bibitem{saarela2016}
Olli Saarela, Leo Belzile, and David~A Stephens.
\newblock A bayesian view of doubly robust causal inference.
\newblock {\em Biometrika}, 103:667--681, 2016.

\bibitem{saarela2015}
Olli Saarela, David~A Stephens, Erica E~M Moodie, and Marina~B Klein.
\newblock On bayesian estimation of marginal structural models.
\newblock {\em Biometrics}, 71:279--301, 2015.

\bibitem{schneeweiss2009}
Sebastian Schneeweiss, Jeremy~A Rassen, Robert~J Glynn, Jerry Avorn, Helen
  Mogun, and M~Alan Brookhart.
\newblock High-dimensional propensity score adjustment in studies of treatment
  effects using health care claims data.
\newblock {\em Epidemiology}, 20:512--522, 2009.

\bibitem{stan}
Stan~Development Team.
\newblock Stan modeling language users guide and reference manual, version
  2.14.0.
\newblock 2016.

\bibitem{vanderlaan2011}
Mark~J. van~der Laan and Sherri Rose.
\newblock {\em Targeted Learning: Causal Inference for Observational and
  Experimental Data}.
\newblock Springer, 2011.

\bibitem{venkitachalam2011}
Lakshmi Venkitachalam, Yang Lei, Elizabeth~A Magnuson, Paul~S Chan, Joshua~M
  Stolker, Kevin~F Kennedy, Neal~S Kleiman, and David~J Cohen.
\newblock Survival benefit with drug-eluting stents in observational studies:
  Fact or artifact?
\newblock {\em Circulation}, 4:587--594, 2011.

\bibitem{wang2012}
Chi Wang, Giovanni Parmigiani, and Francesca Dominici.
\newblock Bayesian effect estimation accounting for adjustment uncertainty.
\newblock {\em Biometrics}, 68:661--686, 2012.

\bibitem{zigler2013}
Corwin~M Zigler, Krista Watts, Robert~W Yeh, Yun Wang, Brent~A Coull, and
  Francesca Dominici.
\newblock Model feedback in bayesian propensity score estimation.
\newblock {\em Biometrics}, 69:263--273, 2013.

\bibitem{zigler2014}
Corwin~Matthew Zigler and Francesca Dominici.
\newblock Uncertainty in propensity score estimation: Bayesian methods for
  variable selection and model-averaged causal effects.
\newblock {\em Journal of the American Statistical Association}, 109:95--107,
  2014.

\end{thebibliography}

\appendix

	\section{Bayesian Propensity Score Algorithm}
	
	\indent
	This section provides further detail on computing the ATE using a Bayesian outcome model as the second step, introduced in the main paper. 
	
	In practice, we first estimate the propensity score model, simulating from the posterior distribution of $\boldsymbol \pi$ using MCMC. To draw samples from our regularized models, we used Hamiltonian Monte Carlo implemented in Stan and analyzed using the RStan package \cite{rstan}\cite{stanjstatsoft}. BART models for the propensity score were fit using the BayesTree package developed by the original authors and available in R \cite{bart}. A regularized logistic model gives draws of the coefficient vector $ \boldsymbol{\widehat{\beta}}_k $, from which we can calculate $\boldsymbol{\widehat{\pi}}_k({\mathbf C})$ as the vector of fitted values: $\boldsymbol{\widehat{\pi}}_k({\mathbf C}) = \mbox{logit}^{-1}  \boldsymbol{\widehat{\beta}}_k  \mathbf{C}$, where we have augmented $\mathbf{C}$ with a vector of 1s as the first column to represent the intercept. Similarly, BART outputs $\boldsymbol{\widehat{\pi}}_k({\mathbf C})$ as fitted value draws from the probit sum-of-trees model.
	
	We now present two different approaches depending on whether we want to integrate over or fix the propensity score. Throughout this section, $i \in \{1,...N\}$ indexes subjects, $k \in \{1,...K\}$ indexes draws from the propensity score model, and $j \in \{1,...J\}$ indexes draws from the outcome model conditional on the propensity score. 
	
	\subsection{Fixed Propensity Score}
	
	Saarela et al (2015) fixed the propensity score to it's posterior mean in analogy to frequentist propensity score analyses, in which the scores are fixed to their maximum likelihood estimate \cite{saarela2015}. The posterior mean propensity score for each subject can simply be calculated as $\mathbb{E}[\widehat{\pi}(\boldsymbol{C}_i)] = \frac{1}{K} \sum_{k = 1}^{K} \widehat{\pi}_k(\boldsymbol{C}_i)$. Then conditional on the vector of mean propensity scores $\mathbb{E}[\boldsymbol{\widehat{\pi}}({\mathbf C})] = \{\mathbb{E}[\widehat{\pi}(\boldsymbol{C}_1)], \mathbb{E}[\widehat{\pi}(\boldsymbol{C}_2)], ... \mathbb{E}[\widehat{\pi}(\boldsymbol{C}_N)]\}$ and the observed data ${\mathbf D} = \{\boldsymbol{Y}, \boldsymbol{X}\}$, we can sample from the closed form beta posterior distributions for $p_1$ and $p_0$:
	
	\begin{align*}
	\widehat{p}_0 \mid \mathbf{D}, \mathbb{E}[\boldsymbol{\widehat{\pi}}({\mathbf C})] &\sim \mbox{Beta}  \big(a_0, b_0 \big)\\
	\widehat{p}_1 \mid \mathbf{D}, \mathbb{E}[\boldsymbol{\widehat{\pi}}({\mathbf C})] &\sim \mbox{Beta}  \big(a_1, b_1 \big)
	\end{align*}
	
	\begin{alignat*}{2}
	&a_1 = \alpha_{11} +  \gamma_1 \bigg(\sum_{i=1}^n \frac{X_i Y_i}{\mathbb{E}[\widehat{\pi}(\boldsymbol{C}_i)] }\bigg) ; ~ &&b_1 = \alpha_{10} + \gamma_1 \bigg(\sum_{i=1}^n  \frac{X_i (1-Y_i)}{\mathbb{E}[\widehat{\pi}(\boldsymbol{C}_i)] } \bigg) ,\\
	&a_0 = \alpha_{00} + \gamma_0 \bigg(\sum_{i=1}^n \frac{(1-X_i) Y_i}{1-\mathbb{E}[\widehat{\pi}(\boldsymbol{C}_i)] } \bigg) ; ~ &&b_0 = \alpha_{01} + \gamma_0 \bigg(\sum_{i=1}^n \frac{(1-X_i) (1-Y_i)}{1-\mathbb{E}[\widehat{\pi}(\boldsymbol{C}_i)] } \bigg).\\
	&\gamma_1 = \frac{\sum_{i=1}^n X_i}{\sum_{i=1}^n X_i / \mathbb{E}[\widehat{\pi}(\boldsymbol{C}_i)] }; ~ &&\gamma_0 = \frac{\sum_{i=1}^n (1-X_i)}{\sum_{i=1}^n (1-X_i) / (1-\mathbb{E}[\widehat{\pi}(\boldsymbol{C}_i)] )}.
	\end{alignat*}

	These can easily be done in R by simulating using the function \texttt{rbeta}. For a single draw $\widehat{p}_{0j}$ and $\widehat{p}_{1j}$, we get a single draw $\widehat{\Delta}_{j}$ from the posterior $\widehat{\Delta} \mid {\mathbf D}, \mathbb{E}[\boldsymbol{\widehat{\pi}}({\mathbf C})]$ as the difference:
	
	$$\widehat{\Delta}_{j} = \widehat{p}_{j1} - \widehat{p}_{j0}$$
	
	We make $J$ such draws to get a representative sample from $\widehat{\Delta} \mid {\mathbf D}, \mathbb{E}[\boldsymbol{\widehat{\pi}}({\mathbf C})]$. The posterior mean causal effect is $\frac{1}{J} \sum_{j=1}^{J} \widehat{\Delta}_j$ with variance $\frac{1}{J-1} \sum_{j=1}^{J} (\widehat{\Delta}_j - \frac{1}{J} \sum_{j=1}^{J} \widehat{\Delta}_j)^2$. Posterior intervals can be calculated from the empirical quantiles of $\{\widehat{\Delta}_1, \widehat{\Delta}_2, ... \widehat{\Delta}_J\}$.
	
	\subsection{Integrating over Propensity Score Distribution}
	
	Kaplan and Chen (2012) and Graham (2016) proposed integrating over the propensity score distribution to propagate uncertainty \cite{kaplan2012}\cite{graham2016}. To integrate over the propensity score distribution we generate draws from the outcome model for each draw from the propensity score model, and base posterior inferences on the resulting $J \times K$ total draws. 
	
	For $k \in \{1,2,...K\}$ we condition on the $k$th propensity score draw $\boldsymbol{\widehat{\pi}}_k({\mathbf C})$ and the data $\mathbf D$ to get draws from the posterior $p_{k1}$ and $p_{k0}$
	
	\begin{align*}
	\widehat{p}_{k0} \mid \mathbf{D}, \boldsymbol{\widehat{\pi}}_k({\mathbf C}) &\sim \mbox{Beta}  \big(a_{k0}, b_{k0} \big)\\
	\widehat{p}_{k1} \mid \mathbf{D}, \boldsymbol{\widehat{\pi}}_k({\mathbf C}) &\sim \mbox{Beta}  \big(a_{k1}, b_{k1} \big)
	\end{align*}
	
	\begin{alignat*}{2}
	&a_{k1} = \alpha_{11} +  \gamma_1 \bigg(\sum_{i=1}^n \frac{X_i Y_i}{\widehat{\pi}_k(\boldsymbol{C}_i)}\bigg) ; ~ &&b_{k1} = \alpha_{10} + \gamma_1 \bigg(\sum_{i=1}^n  \frac{X_i (1-Y_i)}{\widehat{\pi}_k(\boldsymbol{C}_i) } \bigg) ,\\
	&a_{k0} = \alpha_{00} + \gamma_0 \bigg(\sum_{i=1}^n \frac{(1-X_i) Y_i}{1-\widehat{\pi}_k(\boldsymbol{C}_i) } \bigg) ; ~ &&b_{k0} = \alpha_{01} + \gamma_0 \bigg(\sum_{i=1}^n \frac{(1-X_i) (1-Y_i)}{1-\widehat{\pi}_k(\boldsymbol{C}_i) } \bigg).\\
	&\gamma_{k1}= \frac{\sum_{i=1}^n X_i}{\sum_{i=1}^n X_i /\widehat{\pi}_k(\boldsymbol{C}_i) }; ~ &&\gamma_{k0} = \frac{\sum_{i=1}^n (1-X_i)}{\sum_{i=1}^n (1-X_i) / (1-\widehat{\pi}_k(\boldsymbol{C}_i))}.
	\end{alignat*}

	For a single draw $\widehat{p}_{0kj}$ and $\widehat{p}_{1kj}$, we get a single draw $\widehat{\Delta}_{jk}$ from the posterior $\widehat{\Delta} \mid  {\mathbf D}, \boldsymbol{\widehat{\pi}}_k({\mathbf C})$ as the difference:
	
	$$\widehat{\Delta}_{jk} = \widehat{p}_{1kj} - \widehat{p}_{0kj}$$
	
	We make $J$ such draws to get a representative sample from $\widehat{\Delta} \mid {\mathbf D}, \boldsymbol{\widehat{\pi}}_k({\mathbf C})$.
	
	To get the posterior mean and variance of the ATE across both propensity score draws and outcome draws we compute: 
	\begin{align}\label{draws_mean}
	\mathbb{E}(\widehat{\Delta} \mid {\mathbf D}) &= \frac{1}{K}\frac{1}{J} \sum_{k=1}^K \sum_{j=1}^J  \widehat{\Delta}_{jk} \\
	\begin{split}\label{draws_total_variance}
	\mathbb{V}(\widehat{\Delta} \mid {\mathbf D}) ={}& \frac{1}{K} \sum_{k=1}^K \bigg\{ \frac{1}{J-1} \sum_{j=1}^J \bigg(\widehat{\Delta}_{jk} - \frac{1}{J}\sum_{j=1}^J \widehat{\Delta}_{jk}\bigg)^2 \bigg\} \\
	& +  \frac{1}{K-1} \sum_{k=1}^K \bigg\{ \frac{1}{J}\sum_{j=1}^J \widehat{\Delta}_{jk} - \mathbb{E}(\widehat{\Delta} \mid \mathbf{D})  \bigg\}^2
	\end{split}
	\end{align}
	(\ref{draws_mean}) averages over all draws. (\ref{draws_total_variance}) is simply the law of total variance as commonly used for multiple imputation and adds the ``within" and ``between" variance for each propensity score draw. Note that the first term in (\ref{draws_total_variance}) gives the average variance within outcome draws conditional on the propensity score, and the second term in (\ref{draws_total_variance}) gives the variance of the average estimate conditional on the propensity score. 
	
	However we recommend not using (\ref{draws_total_variance}) to get approximate credible intervals, as we can just take the empirical quantiles of the $J \times K$ draws $\widehat{\Delta}_{jk}$ after concatenating across propensity score draws into the vector $\{\Delta_{11}, ...\Delta_{J1},\Delta_{12},..., \Delta_{1K},... \Delta_{JK}\}$. For example, the 2.5th and 97.5th percentiles provide the ends of the 95\% credible interval for $\Delta \mid {\mathbf D}$. Any other posterior summaries (e.g. $\Prob(\Delta > 0)$) can be computed in this way as well.

	\section{Link to Frequentist Inverse Probability Weighting}

	Here we let $\widehat{\boldsymbol \pi} = \widehat{\boldsymbol \pi}({\mathbf C})$ and $\widehat{\pi}(\boldsymbol{C}_i) = \widehat{\pi}_{i}$ for simplicity, we assume that $\widehat{\boldsymbol \pi}$ is correctly estimated, and we ignore its uncertainty. Our goal is just to show that our estimator, 
	
	\begin{equation}
	\widehat{\Delta} \mid {\mathbf D}, \widehat{\boldsymbol \pi} = \widehat{p}_1 - \widehat{p}_0, \label{Y_x}
	\end{equation}
	
	is equal (or arbitrarily close) in expectation to $\widehat{\Delta}_{IPW2}$ from Lunceford and Davidian (2004), i.e. $\mathbb{E}(\widehat{\Delta} \mid {\mathbf D}, \widehat{\boldsymbol \pi}) = \widehat{\Delta}_{IPW2}$, which the authors showed is consistent under the assumption that $\boldsymbol \pi$ is properly specified \cite{lunceford2004}. In fact we only need to show that $\mathbb{E}(p_1) = \widehat{\mu}_1 = \big( \sum_{i=1}^n \frac{X_i}{\pi_i} \big)^{-1} \sum_{i=1}^n \frac{X_i Y_i}{\pi_i}$ and $\mathbb{E}(p_0) = \widehat{\mu}_0 = \big( \sum_{i=1}^n \frac{1-X_i}{1-\pi_i} \big)^{-1} \sum_{i=1}^n \frac{(1-X_i) Y_i}{1-\pi_i}$ due to the linearity of expectations. Note that if $Z \sim \mbox{Beta}(a,b)$, then $\mathbb{E}(Z) = \frac{a}{a+b}$. Assuming improper priors $\alpha_{11} = \alpha_{10} = 0$ for $p_1$ we have: 
	
	\begin{align}
	\mathbb{E}(\widehat{p}_1) &= \frac{\gamma_1 \big( \sum_{i=1}^n \frac{X_i Y_i}{\widehat{\pi}_i} \big)}{\gamma_1 \big( \sum_{i=1}^n \frac{X_i Y_i}{\widehat{\pi}_i} \big) + \gamma_1 \big( \sum_{i=1}^n \frac{X_i (1-Y_i)}{\widehat{\pi}_i} \big)} \\
	&= \frac{\gamma_1}{\gamma_1} \frac{\big( \sum_{i=1}^n \frac{X_i Y_i}{\widehat{\pi}_i} \big)}{\big( \sum_{i=1}^n \frac{X_i Y_i}{\widehat{\pi}_i} \big) + \big( \sum_{i=1}^n \frac{X_i (1-Y_i)}{\widehat{\pi}_i} \big)} \\
	&= \frac{\big( \sum_{i=1}^n \frac{X_i Y_i}{\widehat{\pi}_i} \big)}{\big( \sum_{i=1}^n \frac{X_i Y_i}{\widehat{\pi}_i} \big) + \big( \sum_{i=1}^n \frac{X_i (1-Y_i)}{\widehat{\pi}_i} \big)} \\
	&= \frac{\big( \sum_{i=1}^n \frac{X_i Y_i}{\widehat{\pi}_i} \big)}{\big( \sum_{i=1}^n \frac{X_i Y_i + X_i (1-Y_i)}{\widehat{\pi}_i} \big)}\\
	&= \frac{\big( \sum_{i=1}^n \frac{X_i Y_i}{\widehat{\pi}_i} \big)}{\big( \sum_{i=1}^n \frac{X_i}{\widehat{\pi}_i} \big)}\\
	&= \widehat{\mu}_1
	\end{align}
	
	The arithmetic for $\mathbb{E}(\widehat{p}_0) = \widehat{\mu}_0$ is nearly identical and we omit the details. Thus $\mathbb{E}(\widehat{\Delta} \mid {\mathbf D}, \widehat{\boldsymbol \pi}) = \mathbb{E}(\widehat{p}_1) - \mathbb{E}(\widehat{p}_0) = \widehat{\mu}_1 - \widehat{\mu}_0 = \widehat{\Delta}_{IPW2}$, which is consistent. With proper priors $\mathbb{E}(\widehat{p}_1)$ is not exactly equal to $\widehat{\mu}_1$, but converges to it as $n$ gets large and the data overwhelm the prior. For flat priors $\alpha = 1$ or Jeffrey's priors $\alpha = \frac{1}{2}$, the prior plays very little role for even moderate $n$.
	
	The variance of the difference of two Betas does not have a simple closed form and we use MCMC for inference. We found good coverage of 95\% posterior intervals in our simulations, though other methods like TMLE may be more efficient.

	\section{BART Details}
	
	BART is a nonparametric modeling technique that translates decision tree-based ensemble methods, such as random forests, to a Bayesian framework. Such approaches are especially desirable when the form of the model is unknown, as nonlinearities and interactions in regression problems are accommodated. Chipman et al provide a thorough introduction to the method, and the version of BART we use here is identical to the authors' original BART-probit formulation for binary regression \cite{chipman2010}. Briefly, let $T$ denote a classification tree with it's associated binary decision rules, $M = \{\mu_1, \mu_2, ... \mu_L\}$ a set of parameter values for each of $L$ terminal nodes, $R$ the total number of trees, and $g(\boldsymbol C_i | T, M)$ a function assigning value $\mu_l$ to $\boldsymbol C_i$ (i.e. a single tree model). Then the BART-probit sum-of-trees model for a binary treatment is given by: 
	
	\begin{align}
	\Prob(X_i = 1 | \boldsymbol C_i)  =  \Phi \bigg\{ \sum_{r=1}^R g(\boldsymbol C_i | T_r, M_r) \bigg \},
	\end{align}  
	
	where $\Phi$ is the CDF of the standard normal. Regularizing priors are then placed over $T_r$ to limit the ``depth" of individual trees and over $M_r$ to shrink $\mu_{lr}$ towards zero, controlling the effect of individual terminal nodes. The original authors provide default specifications for these parameters, which we use throughout this paper \cite{chipman2010}.
	
	Specifically, independence is assumed between trees $T_r$ given nodes $M_r$, and between the terminal nodes $\mu_{lr}$ given $T_r$. Thus: 
	
	\begin{align}
	\Prob(T_1, M_1, T_2, M_2, ..., T_R, M_R) &= \prod_{r=1}^{R} \Prob(T_r, M_r) = \prod_{r=1}^R \Prob(M_r | T_r) \Prob(T_r)\\
	\Prob(M_r | T_r) &= \prod_{l=1}^L \Prob(\mu_{lr} | T_r)
	\end{align} 
	
	This greatly simplifies prior specification as we can set a prior relatively intuitively for the individual pieces $\Prob(T_r)$ and $\Prob(\mu_{lr} | T_r)$. 
	
	$\Prob(T_r)$ is specified in 3 parts: the probability that each node is non-terminal (a leaf), the distribution over splitting variables, and the distribution over possible splitting values. The latter two are simply set to be uniform over all potential variables and splitting values. The prior on node depth $d$ is more consequential and is given by: 
	\begin{equation}
	a (1 + d)^{-b},  ~~~\mbox{where}~ a \in (0,1), b \in [0,\infty)
	\end{equation}
	
	$a$ and $b$ are essentially regularizing parameters that control how deep individual trees are allowed to grow. Shallow trees are less flexible and tend towards an additive model (no interactions) at the limit of 1 split.  The authors experimented with various settings to control the depth and recommend the choices $b = 2$ and $a = .95$ as good defaults, which put most prior mass on trees of around depth 2 or 3 \cite{chipman2010}. We used only these settings in our simulated and applied analyses, and did not ``tune" these parameters.
	
	$\Prob(\mu_{lr} | T_r)$ is specified as a normal distribution $\mathcal{N}(0, \sigma_\mu)$ with $\sigma_\mu = 3 / c \sqrt{R}$. This shrinks the parameter estimates towards 0 on the probit scale (i.e. 0.5 on the probability scale). The idea is to put high prior probability that $\Prob(X_i  = 1 | \boldsymbol C_i)$ is in the interval $[\Phi(-3), \Phi(3)] = [.001, .999]$ and the tuning parameter $c$ is chosen to achieve this. The authors recommend $c \in [1,3]$ and suggest $c = 2$ as a good default, which is the specification that we used throughout this article \cite{chipman2010}.
	
	As a final specification, the authors recommend fixing $R$, the size of the forest grown \cite{chipman2010}. The reason for fixing this parameter is to greatly reduce the computational burden that would be incurred by treating $R$ as an unknown parameter.
	
	Gibbs sampling can be used to obtain posterior draws from a BART-probit model. We use the R package \texttt{BayesTree} to fit BART-probit and extract a propensity score distribution \cite{bart}.

	\section{Sparse Simulations}
	
	We repeat the ``realistic" simulation study done in the main paper, but with sparse underlying coefficients. To achieve this, we set 90 of the 100 propensity score coefficients exactly equal to zero. The remaining coefficients had magnitudes between .8 and 1.1, making them relatively strong predictors of treatment and important confounders. Otherwise the model was specified in exactly the same way as in the main paper.
	
	Again, all of our methods achieved a bias reduction compared to the naive estimate. The variance of the $t_3(0,2.5)$ was large and resulted in terrible mean squared error (MSE) and wide confidence intervals that overcovered the true effect. Along with IPW and integrated BART, the MSE was actually worse than an unadjusted estimate, though coverage was improved. 
	
	Integrated Bayesian propensity scores using a logistic treatment model with horseshoe priors performed the best, with tight confidence intervals, low MSE, and coverage very near 95\%. BART had wider confidence intervals and also slightly overcovered the causal effect. All of Bayesian methods had better coverage when integrating over the propensity score, though mean squared error was better when the propensity score was fixed due to the decrease in variance. TMLE is also unbiased and very efficient with the lowest MSE of any method, but suffered from significant undercoverage.

	\begin{table}
		\centering
		\caption{\footnotesize{\textit{Bias (estimate - true effect), confidence interval width, and coverage for various estimators over 500 simulations with sparse coefficients as described above and sample size $n = 1000$.}}}
		\begin{tabular}{lrrrr}
			
			& Bias &  MSE $\times 10^3$& CI Width & Coverage \\ 
			\hline
			Naive Estimate & .013 & 0.95 & .101 & 89.0\% \\
			\hline
			\textit{Integrated Propensity Score} & & & &  \\
			\hline
			Logistic $t_3(0,2.5)$ Priors & -.009 & 3.10 & .234 & 95.6\% \\
			Logistic Horseshoe Priors & .002 & 0.94 & .126 & 95.2\% \\
			BART-Probit & .001 & 1.01 & .203 & 98.8\% \\
			\hline
			\textit{Mean Propensity Score} & & & &  \\
			\hline
			Logistic $t_3(0,2.5)$ Priors& -.002 & 1.74 & .100 & 79.8\% \\
			Logistic Horseshoe Priors & .004 & 0.86 & .100 & 91.6\% \\
			BART-Probit & .005 & 0.71 & .100 & 93.6\% \\
			\hline
			\textit{Frequentist Methods} & & & & \\
			\hline
			Inverse Probability Weighting & -.003 & 1.62 & .164 & 96.8\% \\
			TMLE & .000 & 0.63 & .070 & 84.0\% \\
			\hline
		\end{tabular}
		\label{sparse_simulation_table}
	\end{table}

	\section{Complete Mass-DAC Covariates}
	The full list of covariates from the Mass-DAC clinical registry appears in table \ref{tab:mass_dac_covariates}. We include prevalences or mean and standard deviation overall and in each treatment group. For the sake of space and to conform to privacy requirements, we summarize the site and artery segment variables for which there are 25 and 31 categories, respectively.

	{\footnotesize	
		\begin{longtable}{lccc}
			
			\caption{\footnotesize{\textit{Prevalence or mean (SD) of variables in Mass-DAC registry. Columns denote prevalence or mean (SD) overall and conditional on treatment received. All potential confounders are pre-treatment. Cells representing data on fewer than 10 patients are suppressed in accordance with privacy guidelines. TVR = target vessel revascularization; AMI = acute myocardial infarction; HMO = health maintenance organization; CAD = coronary artery disease; CVD = cardiovascular disease; PAD = peripheral artery disease; CABG = coronary artery bypass graft; STEMI = ST-elevated myocardial infarction; LVSD = left ventricular systolic dysfunction; NYHA = New York heart association; LAD = left anterior descending (artery); RCA = right coronary artery.}}}\\
			\hline
			\textit{\textbf{Registry Outcomes}} & Overall (n = 8718) & BMS (n = 3081) & DES (n = 5637)\\ 
			\hline
			1 Year Mortality-AMI-TVR Composite & 17.96 & 24.02 & 14.65 \\  
			1 Year Mortality & 5.75 & 10.19 & 3.32 \\ 
			30 Day Mortality & 2.01 & 4.19 & 0.82 \\ 
			
			\hline
			\textit{\textbf{Registry Confounders}} \\
			Male & 69.39 & 68.16 & 70.05 \\ 
			Mean Age (SD) & 64.66 (12.5) & 66.37 (11.7) & 63.72 (13.5) \\ 
			Mean Height (SD) & 170.67 (10.5) & 170.41 (10.6) & 170.81 (10.4) \\ 
			Mean Weight (SD) & 86.48 (20.3) & 84.89 (19.9) & 87.34 (20.8) \\
			
			\rowgroup{Race}\\
			\MyIndent Caucasian & 91.26 & 90.72 & 91.56 \\ 
			\MyIndent Black & 3.79 & 4.87 & 3.19 \\ 
			\MyIndent Asian & 2.31 & 2.43 & 2.24 \\ 
			\MyIndent Native American & $<0.1$ & $<0.3$& $<0.2$ \\ 
			\MyIndent Native Pacific & $<0.1$& $<0.3$& $<0.2$\\ 
			
			Hispanic or Latino  & 4.55 & 4.8 & 4.42 \\ 
			
			\rowgroup{Payor}\\
			\MyIndent Government & 51.57 & 59.17 & 47.42\\
			\MyIndent None & 2.56 & 4.12 & 1.7 \\ 
			\MyIndent Non-US & $<0.1$ & $<0.3$ & $<0.2$ \\ 
			\MyIndent Private Commercial or HMO & 45.82 & 36.68 & 50.82 \\ 
			
			Smoker & 74.68 & 70.85 & 76.78 \\ 
			Hypertension & 78.3 & 76.6 & 79.23 \\ 
			Dyslipidemia & 81.82 & 77.96 & 83.93 \\ 
			Diabetes & 32.16 & 30.38 & 33.14 \\ 
			Family History CAD & 28.96 & 24.08 & 31.63 \\ 
			Chronic Lung Disease & 14.21 & 16.42 & 13 \\ 
			Current Dialysis & 1.82 & 2.21 & 1.61 \\ 
			
			Prior CVD & 10.23 & 11.33 & 9.63 \\ 
			Prior PAD & 11.96 & 12.76 & 11.53 \\ 
			Prior Myocardial Infarction & 26.67 & 25.84 & 27.12 \\ 
			Prior Heart Failure & 10.61 & 12.46 & 9.6 \\ 
			Prior Valve Surgery & 1.56 & 2.34 & 1.14 \\ 
			Prior PCI & 27.56 & 19.77 & 31.83 \\ 
			Prior CABG & 12.3 & 11.46 & 12.76 \\ 
			Prior Cardiogenic Shock & 1.9 & 3.89 & 0.82 \\ 
			Prior Cardiac Arrest & 2.29 & 4.15 & 1.28 \\ 
			
			\rowgroup{CAD Presentation}\\
			\MyIndent No Angina & 5.32 & 5.78 & 5.07 \\ 
			\MyIndent Symptom Unlikely to be Ischemic & 1.00 & 1.17 & 0.90 \\ 
			\MyIndent Stable Angina & 11.88 & 5.81 & 15.2 \\ 
			\MyIndent Unstable Angina & 30.19 & 23.4 & 33.9 \\ 
			\MyIndent Non-STEMI & 27.23 & 28.17 & 26.72 \\ 
			\MyIndent STEMI & 24.37 & 35.67 & 18.2 \\ 
			
			Thrombolytic Therapy & 0.8 & 1.23 & 0.57 \\ 
			Cardiomyopathy or LVSD & 9.36 & 11.1 & 8.41 \\ 
			
			\rowgroup{Anginal Canadian Classification}\\
			\MyIndent 0 & 9.64 & 11.65 & 8.53\\
			\MyIndent I & 2.4 & 1.59 & 2.84 \\ 
			\MyIndent II & 12.2 & 7.66 & 14.69 \\ 
			\MyIndent III & 28.83 & 24.15 & 31.38 \\ 
			\MyIndent IV & 46.94 & 54.95 & 42.56 \\ 
			
			\rowgroup{Anti-Anginal Medications}\\
			\MyIndent Beta Blockers & 57.79 & 54.56 & 59.55 \\ 
			\MyIndent Calcium Channel Blockers & 11.44 & 11.39 & 11.46 \\ 
			\MyIndent Long Acting Nitrates & 11.49 & 9.41 & 12.63 \\ 
			\MyIndent Ranolazine & 0.62 & 0.45 & 0.71 \\ 
			\MyIndent Other Agent & 1.4 & 1.01 & 1.61 \\ 
			
			\rowgroup{NYHA Class}\\
			\MyIndent 0 & 87.57 & 83.51 & 89.78\\
			\MyIndent I & 0.91 & 0.78 & 0.98 \\ 
			\MyIndent II & 3.06 & 3.67 & 2.73 \\ 
			\MyIndent III & 4.16 & 5.39 & 3.49 \\ 
			\MyIndent IV & 4.3 & 6.65 & 3.02 \\

			Compassionate Use & 0.91 & 2.01 & 0.3 \\ 
			Cardiogenic Shock & 1.85 & 3.8 & 0.78 \\ 
			Mechanical Ventricular Support & 0.58 & 1.14 & 0.28 \\ 
			Ejection Fraction $<30\%$ & 2.88 & 3.7 & 2.43 \\ 
			
			\rowgroup{Coronary Anatomy}\\
			\MyIndent Left Dominant & 8.30 & 8.08 & 8.42 \\
			\MyIndent Right Dominant & 86.19 & 85.88 & 86.36 \\ 
			\MyIndent Left Dominant & 5.51 & 6.04 & 5.22 \\ 
			
			Left Main Disease & 5.93 & 6.52 & 5.61 \\ 
			Mean Left Main Stenosis (SD) & 7.98 (19.8) & 8.74(19.3) & 7.57(20.7) \\ 
			Mean Proximal LAD Stenosis (SD)& 36 (39.7) & 35.1(39.7) & 36.5(39.5) \\ 
			Mean Mid-Distal LAD Stenosis (SD)& 48.2 (39.7) & 47.64(39.7) & 48.51(39.6) \\ 
			Mean Circumflex Stenosis (SD)& 47.95 (40.8) & 47.94(40.7) & 47.96(40.8) \\ 
			Mean RCA Stenosis (SD)& 60.66 (39.9) & 65.25(40) & 58.15(39.2) \\ 
			
			\rowgroup{Status}\\
			\MyIndent Urgent & 52.49 & 48.78 & 54.51 \\ 
			\MyIndent Emergent & 26.69 & 38.30 & 20.35 \\ 
			\MyIndent Other & 20.82 & 12.92 & 25.14\\
			
			\rowgroup{PCI Indication}\\
			\MyIndent Immediate PCI for STEMI & 21.80 & 31.39 & 16.55\\
			\MyIndent STEMI (Unstable, $>12$ hours) & 1.67 & 2.66 & 1.14 \\ 
			\MyIndent STEMI (Stable, $>12$ hours) & 0.57 & 0.94 & 0.37 \\ 
			\MyIndent STEMI (Stable, thrombolytics) & 0.44 & 0.58 & 0.35 \\ 
			\MyIndent STEMI (Rescue, failed thrombolytics) & 0.46 & 0.75 & 0.3 \\ 
			\MyIndent High risk Non-STEMI & 48.99 & 45.86 & 50.7 \\ 
			\MyIndent Staged & 0.86 & 0.45 & 1.08 \\ 
			\MyIndent Other & 25.21 & 17.36 & 29.5 \\ 
			
			Thrombectomies Used & 0.12 (0.3) & 0.16(0.3) & 0.09(0.4) \\ 
			Lesions Treated & 1.3 (0.6) & 1.24 (0.6) & 1.33 (0.5) \\
			Lesion Length & 18.08 (10) & 17 (10.4) & 18.68 (9.3) \\ 
			Chronic Total Occlusion & 1.73 & 1.3 & 1.97 \\ 
			In-stent Restenosis & 6.96 & 2.34 & 9.49 \\ 
			Total Stents Used & 1.46 (0.8) & 1.41 (0.8) & 1.5 (0.8) \\ 
			
			\rowgroup{Sites (25 Total)}\\
			\MyIndent  Min Volume & 47 & 12 & 14\\
			\MyIndent  Mean Volume (SD) & 349 (269) & 123 (99) & 225 (187)\\
			\MyIndent  Max Volume & 896 & 352 & 605 \\

			\rowgroup{Coronary Artery Segments (31 Total)}\\
			\MyIndent Min Volume & 2 & 1 & 1 \\ 
			\MyIndent Mean Volume (SD) & 375 (534) & 126 (183) & 249 (358) \\
			\MyIndent Max Volume & 1903 & 611 & 1335 \\
			
			\hline
			\label{tab:mass_dac_covariates} 
		\end{longtable}
	}
	\newpage
	
	{\footnotesize
		\begin{longtable}{lccc}
			\caption{Prevalence of top 30 most frequently recorded ICD-9 diagnosis codes in CHIA billing data, overall and by treatment group. Codes are grouped into 3 digit broad categories for simplicity of presentation while models use more-granular 5 digit codes.}\\
			\hline
			ICD-9 Code & Overall & BMS & DES \\ 
			\hline
			414: Other forms of chronic ischemic heart disease & 86.24 & 88.28 & 85.12 \\ 
			272: Disorders of lipoid metabolism & 69.47 & 62.67 & 73.18 \\ 
			401: Essential hypertension & 62.20 & 57.25 & 64.91 \\ 
			410: Acute myocardial infarction & 53.99 & 66.44 & 47.19 \\ 
			250: Diabetes mellitus & 30.36 & 27.98 & 31.67 \\ 
			305: Nondependent abuse of drugs & 21.51 & 25.45 & 19.35 \\ 
			411: Other acute and subacute form of ischemic heart disease & 19.65 & 15.74 & 21.78 \\ 
			427: Cardiac dysrhythmias & 17.13 & 24.41 & 13.15 \\ 
			530: Diseases of esophagus & 16.87 & 15.97 & 17.37 \\ 
			428: Heart failure & 15.46 & 20.12 & 12.91 \\ 
			278: Obesity and other hyperalimentation & 12.03 & 11.13 & 12.52 \\ 
			585: Chronic renal failure & 9.77 & 11.33 & 8.92 \\ 
			403: Hypertensive renal disease & 8.99 & 10.52 & 8.16 \\ 
			413: Angina pectoris & 7.90 & 4.35 & 9.85 \\ 
			244: Acquired hypothyroidism & 7.26 & 7.27 & 7.26 \\ 
			496: Chronic airways obstruction, not elsewhere classified & 7.25 & 9.12 & 6.23 \\ 
			300: Neurotic disorders & 6.61 & 6.07 & 6.90 \\ 
			424: Other diseases of endocardium & 6.06 & 8.34 & 4.81 \\ 
			443: Other peripheral vascular disease & 5.38 & 5.55 & 5.29 \\ 
			285: Other and unspecified anemias & 5.23 & 7.59 & 3.94 \\ 
			426: Conduction disorders & 4.90 & 6.78 & 3.87 \\ 
			996: Complications peculiar to certain specified procedures & 4.87 & 2.79 & 6.01 \\ 
			493: Asthma & 4.84 & 5.42 & 4.52 \\ 
			600: Hyperplasia of prostate & 4.82 & 5.52 & 4.43 \\ 
			327: Organic sleep disorders & 4.76 & 4.06 & 5.14 \\ 
			311: Depressive disorder, not elsewhere classified & 4.67 & 5.00 & 4.49 \\ 
			276: Disorders of fluid, electrolyte, and acid-base balance & 4.40 & 6.17 & 3.44 \\ 
			715: Osteoarthrosis and allied disorders & 3.46 & 3.51 & 3.44 \\ 
			724: Other and unspecified disorders of back & 3.30 & 3.64 & 3.12 \\ 
			274: Gout & 3.27 & 3.34 & 3.23\\
			\hline
			\label{tab:poa_codes}  
		\end{longtable}
	}

\end{document}